\documentclass[12pt,preprint]{aastex}
\pdfoutput=1

\def\dnu{$\Delta\nu$}
\def\numax{$\nu_{\rm max}$}
\def\teff{$T_{\rm eff}$}

\def\feh{[Fe/H]}

\shorttitle{Asteroseismic fundamental properties of solar-type
  stars}\shortauthors{Chaplin et al.}  \shortauthors{Chaplin et al.}

\begin{document}

\title{Asteroseismic fundamental properties of solar-type stars
  observed by the NASA \emph{Kepler} Mission}

\author{
   W.~J.~Chaplin\altaffilmark{1,2},
   S.~Basu\altaffilmark{3},
   D.~Huber\altaffilmark{4,5},
   A~Serenelli\altaffilmark{6},
   L.~Casagrande\altaffilmark{7},
   V.~Silva~Aguirre\altaffilmark{2},
   W.~H.~Ball\altaffilmark{8,9},
   O.~L.~Creevey\altaffilmark{10,11},
   L.~Gizon\altaffilmark{9,8},
   R.~Handberg\altaffilmark{1,2},
   C.~Karoff\altaffilmark{2},
   R.~Lutz\altaffilmark{8,9},
   J.~P.~Marques\altaffilmark{8,9},
   A.~Miglio\altaffilmark{1,2},
   D.~Stello\altaffilmark{12,2},
   M.~D.~Suran\altaffilmark{13},
   D.~Pricopi\altaffilmark{13},
   T.~S.~Metcalfe\altaffilmark{14,2},
   M.~J.~P.~F.~G.~Monteiro\altaffilmark{15},
   J.~Molenda-\.Zakowicz\altaffilmark{16},
   T.~Appourchaux\altaffilmark{11},
   J.~Christensen-Dalsgaard\altaffilmark{2},
   Y.~Elsworth\altaffilmark{1,2},
   R.~A.~Garc\'ia\altaffilmark{17},
   G.~Houdek\altaffilmark{2},
   H.~Kjeldsen\altaffilmark{2},
   A.~Bonanno\altaffilmark{18},
   T.~L.~Campante\altaffilmark{1,2},
   E.~Corsaro\altaffilmark{19,18},
   P.~Gaulme\altaffilmark{20},
   S.~Hekker\altaffilmark{21,9},
   S.~Mathur\altaffilmark{14,22},
   B.~Mosser\altaffilmark{23},
   C.~R\'egulo\altaffilmark{24,25},
   D.~Salabert\altaffilmark{26}
}

\altaffiltext{1}{School of Physics and Astronomy, University of
  Birmingham, Edgbaston, Birmingham, B15 2TT, UK}

\altaffiltext{2}{Stellar Astrophysics Centre (SAC), Department of
  Physics and Astronomy, Aarhus University, Ny Munkegade 120, DK-8000
  Aarhus C, Denmark}

\altaffiltext{3}{Department of Physics and Astronomy, Yale University,
  P.O. Box 208101, New Haven, CT, 06520, USA}

\altaffiltext{4}{NASA Ames Research Center, MS 244-30, Moffett Field,
  CA 94035, USA}

\altaffiltext{5}{NASA Postdoctoral Program Fellow}

\altaffiltext{6}{Instituto de Ci\`encias del Espacio (CSIC-IEEC),
  Facultad de Ci\`encies, Campus UAB, E-08193 Bellaterra, Spain}

\altaffiltext{7}{Research School of Astronomy and Astrophysics, Mount
Stromlo Observatory, The Australian National University, ACT 2611,
Australia}

\altaffiltext{8}{Institut f\"ur Astrophysik, Georg-August-Universit\"at
  G\"ottingen, D-37077 G\"ottingen, Germany}

\altaffiltext{9}{Max-Planck-Institut f\"ur Sonnensystemforschung,
  37191 Katlenburg-Lindau, Germany}

\altaffiltext{10}{Universit\'e de Nice Sophia-Antipolis, Laboratoire
  Lagrange, UMR 7293, CNRS, Observatoire de la C\^ote d'Azur, Nice,
  France}

\altaffiltext{11}{Institut d'Astrophysique Spatiale, Universit\'e Paris
  XI -- CNRS (UMR8617), Batiment 121, 91405 Orsay Cedex, France}

\altaffiltext{12}{Sydney Institute for Astronomy, School of Physics,
  University of Sydney, Sydney, Australia}

\altaffiltext{13}{Astronomical Institute of the Romanian Academy,
  Str. Cutitul de Argint, 5, RO 40557, Bucharest, Romania}

\altaffiltext{14}{Space Science Institute, 4750 Walnut Street Suite
  205, Boulder CO 80301, USA}

\altaffiltext{15}{Centro de Astrof\'\i sica, Universidade do Porto, Rua
  das Estrelas, 4150-762, Portugal}

\altaffiltext{16}{Astronomical Institute, University of Wroc\l{}aw,
  ul. Kopernika, 11, 51-622 Wroc\l{}aw, Poland}

\altaffiltext{17}{Laboratoire AIM, CEA/DSM -- CNRS -- Universit\'e Paris
  Diderot -- IRFU/SAp, 91191 Gif-sur-Yvette Cedex, France}

\altaffiltext{18}{INAF -- Astrophysical Observatory of Catania, Via
S. Sofia 78, I-95123 Catania, Italy}

\altaffiltext{19}{Instituut voor Sterrenkunde, KU Leuven,
Celestijnenlaan 200D, B-3001 Leuven, Belgium}

\altaffiltext{20}{Department of Astronomy, New Mexico State
  University, P.O. Box 30001, MSC 4500, Las Cruces, NM 88003-8001,
  USA}

\altaffiltext{21}{Astronomical Institute ``Anton Pannekoek'', University
  of Amsterdam, Science Park 904, 1098 XH Amsterdam, The Netherlands}

\altaffiltext{22}{High Altitude Observatory, NCAR, P.O. Box 3000,
  Boulder, CO 80307, USA}

\altaffiltext{23}{LESIA, CNRS, Universit\'e Pierre et Marie Curie,
  Universit\'e Denis Diderot, Observatoire de Paris, F-92195 Meudon
  cedex, France}

\altaffiltext{24}{Instituto de Astrof\'isica de Canarias, 38200 La
  Laguna, Tenerife, Spain}

\altaffiltext{25}{Departamento de Astrof\'isica, Universidad de La
  Laguna, 38206 La Laguna, Tenerife, Spain}

\altaffiltext{26}{Laboratoire Lagrange, UMR7293, Universit\'e de Nice
  Sophia-Antipolis, CNRS, Observatoire de la C\^ote d'Azur, Bd. de
  Observatoire, 06304 Nice, France}

\begin{abstract}

We use asteroseismic data obtained by the NASA \emph{Kepler} Mission
to estimate the fundamental properties of more than 500 main-sequence
and sub-giant stars. Data obtained during the first 10\,months of
\emph{Kepler} science operations were used for this work, when these
solar-type targets were observed for one month each in a survey
mode. Stellar properties have been estimated using two global asteroseismic
parameters and complementary photometric and spectroscopic
data. Homogeneous sets of effective temperatures, $T_{\rm eff}$, were
available for the entire ensemble from complementary photometry;
spectroscopic estimates of $T_{\rm eff}$ and [Fe/H] were available
from a homogeneous analysis of ground-based data on a subset of 87
stars. 

We adopt a grid-based analysis, coupling six pipeline codes to eleven
stellar evolutionary grids. Through use of these different
grid-pipeline combinations we allow implicitly for the impact on the
results of stellar model dependencies from commonly used grids, and
differences in adopted pipeline methodologies. By using just two
global parameters as the seismic inputs we are able to perform a
homogenous analysis of all solar-type stars in the asteroseismic
cohort, including many targets for which it would not be possible to
provide robust estimates of individual oscillation frequencies (due to
a combination of low S/N and short dataset lengths). The median final
quoted uncertainties from consolidation of the grid-based analyses are
for the full ensemble (spectroscopic subset) approximately $10.8$\,\%
(5.4\,\%) in mass, 4.4\,\% (2.2\,\%) in radius, 0.017\,dex
(0.010\,dex) in $\log\,g$, and 4.3\,\% (2.8\,\%) in mean
density. Around 36\,\% (57\,\%) of the stars have final age
uncertainties smaller than 1\,Gyr. These ages will be useful for
ensemble studies, but should be treated carefully on a star-by-star
basis.

Future analyses using individual oscillation frequencies will offer
significant improvements on up to 150 stars, in particular for
estimates of the ages, where having the individual frequency data is
most important.

\end{abstract}


\section{Introduction}
\label{sec:intro}

Recent advances in observational asteroseismology are making it
possible to estimate accurate and precise fundamental properties of a
growing number of solar-type stars. These advances have come in large
part from new satellite observations, for example from the French-led
CoRoT satellite (e.g., Michel et al. 2008; Appourchaux et al. 2008;
Michel \& Baglin 2012), and in particular the NASA \emph{Kepler}
Mission (Chaplin et al. 2010; Gilliland et al. 2010a).

During the first 10\,months of science operations more than 2000
solar-type stars were selected by the \emph{Kepler} Asteroseismic
Science Consortium (KASC) to be observed as part of an asteroseismic
survey of the Sun-like population in the \emph{Kepler} field of
view. Solar-like oscillations were detected by \emph{Kepler} in more
than 500 stars (Chaplin et al. 2011), and from these data robust
global or average asteroseismic parameters were determined for all
targets in the sample.  These asteroseismic parameters allow us to
estimate fundamental properties of the stars. In this paper we present
stellar properties---namely masses, radii, surface gravities, mean
densities and ages---of this asteroseismic sample of main-sequence and
subgiant stars.

The most precise asteroseismically derived stellar properties are
obtained when the frequencies of individual modes of oscillation are
modeled (see e.g., Metcalfe et al. 2010, 2012; Silva Aguirre et
al. 2013). Recent noteworthy examples have included several solar-type
exoplanet host stars observed by \emph{Kepler} (e.g., see Batalha et
al. 2011, Howell et al. 2012, Carter et al. 2012, Gilliland et
al. 2013, Chaplin et al. 2013). When the signal-to-noise ratios (S/N)
in the asteroseismic data are too low to allow robust fitting of
individual mode frequencies, or the frequency resolution is
insufficient to resolve clearly the mode structure, it is nevertheless
still possible to extract average or global asteroseismic
parameters. The main parameters are the average large frequency
separation, \dnu, and the frequency of maximum oscillations power,
\numax. Automated analysis codes developed for application to
\emph{Kepler} data (e.g., see Chaplin et al. 2008, Stello et al. 2009)
have enabled efficient extraction of these parameters on large numbers
of stars, even at quite low S/N levels.

Here, the use of global asteroseismic parameters has allowed us to
determine the properties of over 500 stars, rather than the propeties
of just the smaller cohort of around 150 stars observed continuously
over long periods by \emph{Kepler} for which robust frequencies may be
determined.  Detailed studies of several \emph{Kepler} targets have
shown that results obtained using only global parameters provide a
good match (within their uncertainties) to those given by analysis of
individual frequencies (e.g., see exoplanet host-star references
above; also Mathur et al. 2012, Metcalfe et al. 2012, Do\u{g}an et
al. 2013, Silva Aguirre et al. 2013), although fractional
uncertainties in the estimated properties are usually inferior (most
notably the uncertainties in the estimated ages).

We use a grid-based method to determine stellar properties, but with
the powerful diagnostic information contained in the seismic
parameters also brought to bear.  This is the classic approach of
matching the observations to well-sampled grids of stellar
evolutionary models (tracks, or isochrones).  It is not uncommon in
the stellar literature for grid-based estimates of stellar properties
to be presented from analyses that involve one pipeline code coupled
to only one grid of models (with a given input physics). We adopt the
grid-based analysis, but here we couple six pipeline codes to eleven
stellar evolutionary grids. By using a range of grid-pipeline
combinations---comprising a selection of widely used stellar
evolutionary models, covering a range of commonly adopted input
physics---we allow implicitly in our final estimates for the impact of
stellar model dependencies from commonly used grids and physics, and
also differences in adopted analysis pipeline methodologies.

\section{Determination of stellar properties using global asteroseismic parameters}
\label{sec:method}

 \subsection{Global asteroseismic parameters}
 \label{sec:globalast}

\begin{figure*}

\epsscale{0.75}
\plotone{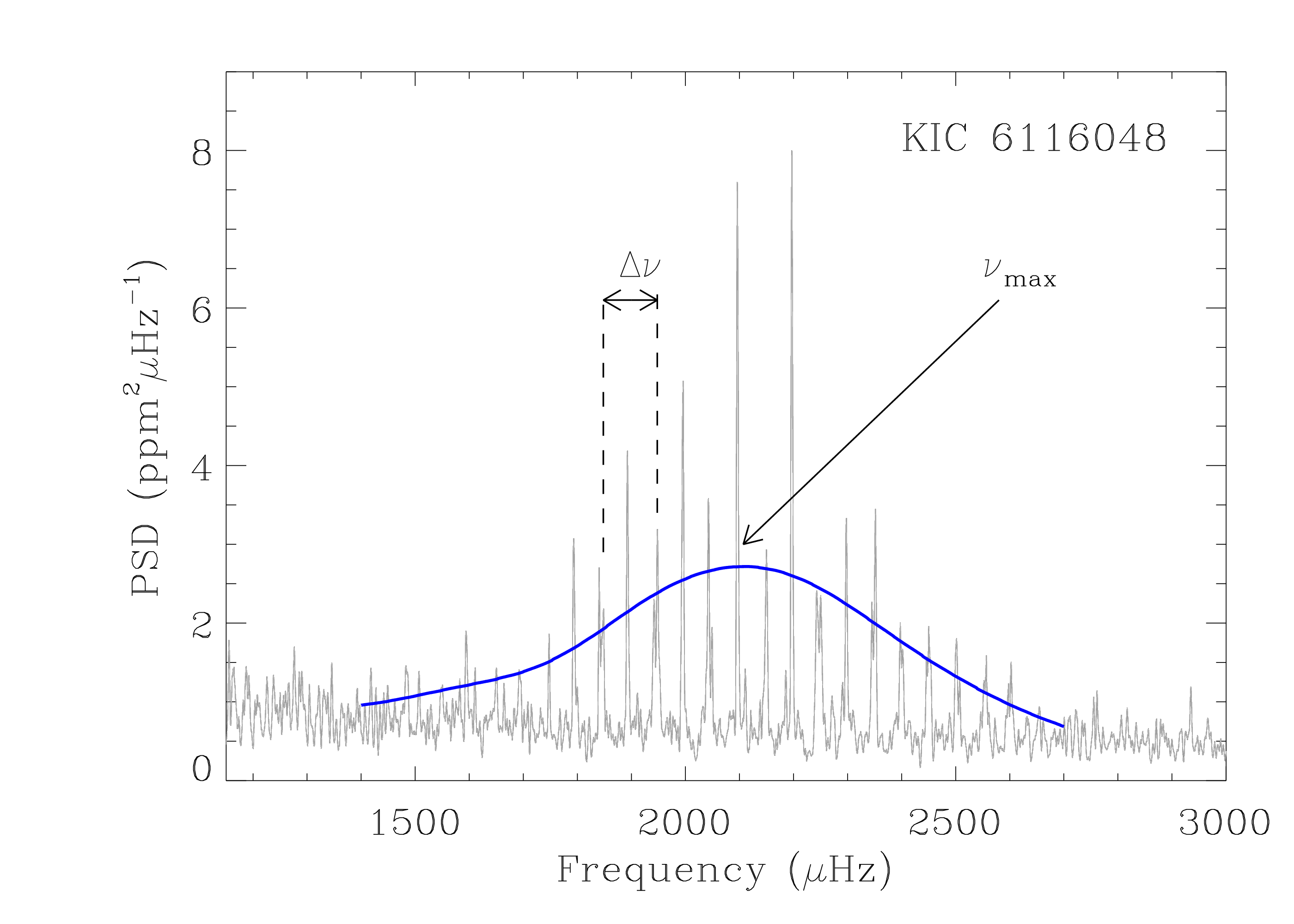}
\epsscale{0.75}
\plotone{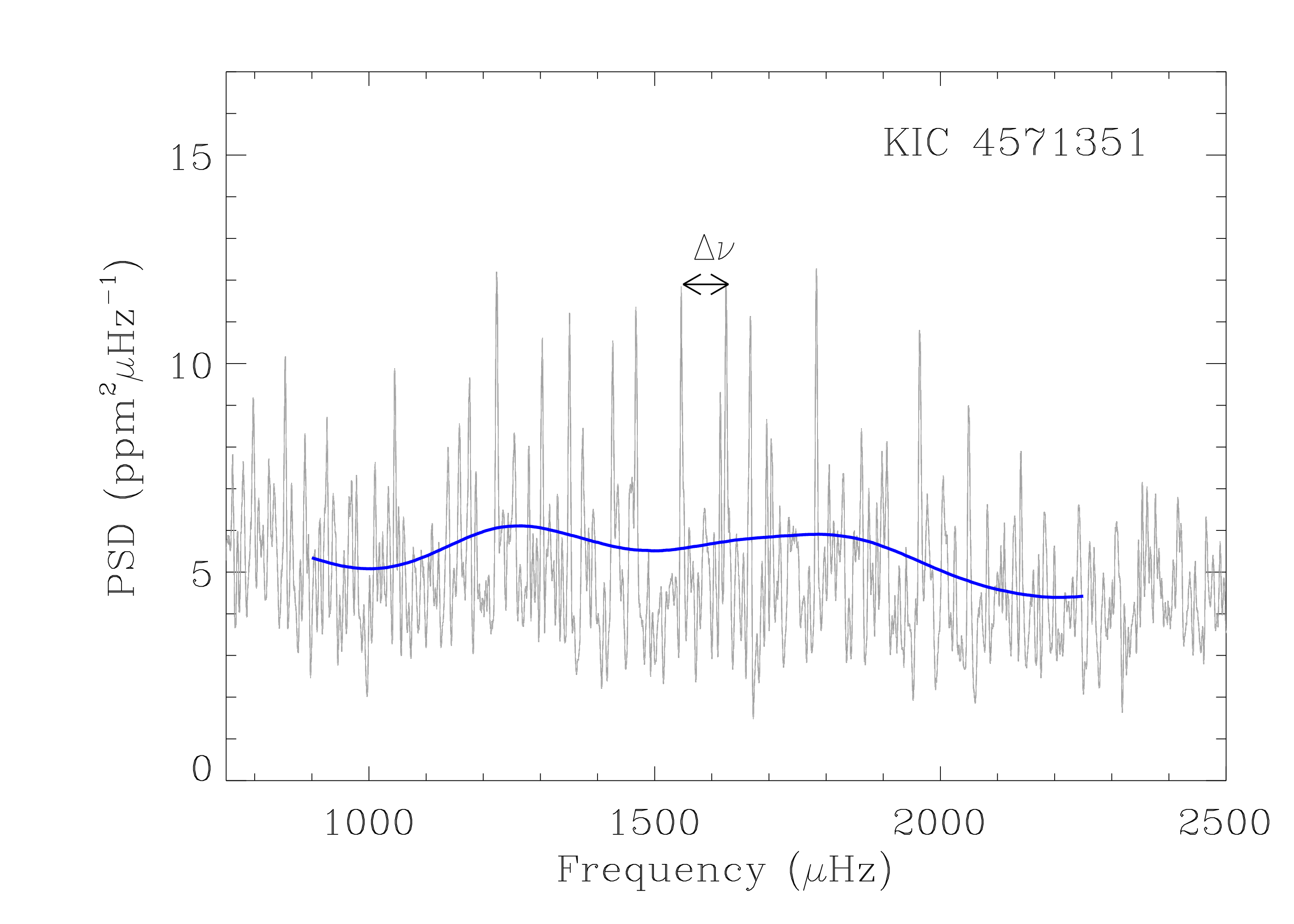}

\caption{Top panel: Frequency power spectrum of KIC\,6116048, computed
  from one month of \emph{Kepler} data. The blue line shows the
  envelope of power given by the oscillations, from heavily smoothing
  the power spectrum (and here multiplied by a factor of five to show
  the envelope more clearly). Bottom panel: Frequency power spectrum
  of the much fainter KIC\,4571351. The higher levels of shot noise
  compared to KIC\,6116048 lead to much lower levels of S/N in the
  oscillations spectrum, making it difficult to extract a robust
  estimate of \numax.}

\label{fig:pspec}
\end{figure*}

Cool subgiants and low-mass, main-sequence stars show rich spectra of
solar-like oscillations, small-amplitude pulsations that are excited
and damped intrinsically by convection in the outer parts of the
star. The most prominent oscillations are acoustic (pressure, or p)
modes of high radial order, $n$.  The observed power in the
oscillations is modulated in frequency by an envelope that typically
has an approximately Gaussian shape (as can be seen in
Figure~\ref{fig:pspec}). The frequency of maximum oscillations power,
\numax, has been shown to scale to good approximation as $gT_{\rm
  eff}^{-1/2}$ (Brown et al. 1991, Kjeldsen \& Bedding 1995, Chaplin
et al. 2008, Belkacem et al. 2011), where $g$ is the surface gravity
and $T_{\rm eff}$ is the effective temperature of the star. The radial
order at \numax\ ranges from $n \simeq 15$ to 19 in sub-giants, and $n
\simeq 17$ to 25 in main-sequence stars. The transition from the
main-sequence to the sub-giant phase occurs at approximately $\nu_{\rm
  max}= 2000\,\rm \mu Hz$ in stars of solar mass and composition; this
frequency decreases to approximately $800\,\rm \mu Hz$ at $M \simeq
\rm 1.5M_{\odot}$.

The most obvious frequency spacings in the spectrum are the large
frequency separations, \dnu, between consecutive overtones $n$ of the
same spherical angular degree, $l$. The average large separation
scales to very good approximation as $\rho^{1/2}$, $\rho \propto
M/R^3$ being the mean density of a star having mass $M$ and surface
radius $R$ (e.g. see Tassoul 1980, Ulrich 1986, Christensen-Dalsgaard
1993).

  \subsection{Principles of stellar properties estimation}
  \label{sec:prin}

As mentioned earlier, we use grid-based methods to determine the
stellar properties, but unlike earlier works we use seismic
observables---here, the global parameters \dnu\ and \numax---together
with non-seismic inputs---here, complementary estimates of $T_{\rm
  eff}$ and the metallicity [Fe/H]---to determine stellar properties.

The information encoded in \dnu\ may be employed in one of two ways.
One may compute theoretical oscillation frequencies of each model in
the grid, and from those frequencies calculate a suitable average
\dnu\ for comparison with the observations, e.g., from $l=0$
(radial-mode) frequencies spanning the same orders $n$ as those
detected in the data. Alternatively, one may circumvent the need to
compute individual oscillation frequencies of every model and instead
use the dependence of \dnu\ on the mean stellar density
(Section~\ref{sec:globalast}) as a scaling relation normalized by solar
properties and parameters, i.e.,
  \begin{equation} 
  \frac{\Delta \nu }{\Delta \nu _{\odot }} \simeq 
       \sqrt{\frac{M/M_{\odot }}{(R/R_{\odot })^{3}}}.
  \label{eq:dnu}
  \end{equation}
The fundamental properties of the models (i.e., $R$ and $M$) are
thence used as inputs to Equation~\ref{eq:dnu} to calculate
model-values of \dnu, against which differences with the observations
may be computed.  Comparison of predictions of Equation~\ref{eq:dnu}
with predictions of \dnu\ from model-calculated eigenfrequencies
reveal small but systematic offsets for solar-type stars, which can be
as large as $\simeq 2\,\%$ (e.g., see Ulrich 1986, White et
al. 2011). When plotted as a function of $T_{\rm eff}$, these
differences manifest as a ``boomerang'' shaped trend (cf. Figures 5
and 6 of White et al. 2011).  The median uncertainty in
\dnu\ in our sample is 2.1\,\%. As we shall see later, this effect is
clearly detectable in our results.

One phenomenon that none of the above allows for is the impact of poor
modelling of the near-surface layers of stars. In the case of the Sun,
this has all been shown to lead to a frequency dependent offset
between observed p-mode frequencies and the model-predicted p-mode
frequencies (e.g., see Christensen-Dalsgaard \& Gough 1984;
Dziembowski et al. 1988; Christensen-Dalsgaard et al. 1997; Kjeldsen
et al. 2008; Chaplin \& Miglio 2013, and references therein), with the
model frequencies being on average too high by a few $\rm \mu Hz$.
This offset, which is sometimes called the ``surface term'', is larger
in modes at higher frequencies. The average large frequency separation
will also be affected by the surface term, by an amount that depends
on the gradient of the frequency offset with radial order, $n$.  In
the case of the Sun, the model-predicted \dnu\ is about 0.75\,\%
higher than the observed \dnu. The impact on grid-search results for
the Sun is relatively small, about 0.3\,\% in the inferred radius, and
not a cause for concern given the level of the uncertainties in the
global asteroseismic parameters used here (e.g., see Basu et
al. 2010). Offsets for other solar-type stars would have to be
substantially larger than the solar offsets to give significant bias
in our results. However, we still await definitive results on the
surface-term offsets, although existing studies suggest that offsets
may be Sun-like in size when stars have close-to-solar surface
properties (e.g., Kjeldsen et al. 2008, Mathur et al. 2012;
Gruberbauer et al. 2013).

Information encoded in \numax\ is currently employed only in
scaling-relation form, i.e., with reference to Section~\ref{sec:globalast}
we have:
  \begin{equation} 
  {{\nu _{\rm max}}\over {\nu _{\rm max,\odot }}}\simeq 
  {{M/M_{\odot }}\over {(R/R_{\odot })^2\sqrt{(T_{\rm eff}/T_{{\rm eff},\odot })}}},
  \label{eq:numax}
  \end{equation}
so that again it is the properties of the models ($R$, $M$, $T_{\rm
  eff}$) that yield model-values of \numax, against which the
observations are compared.

The form of the above equations of course indicates that we do not
necessarily need to employ a grid of stellar evolutionary models: if
\dnu, \numax\ and \teff\ are known for a star, Equations~\ref{eq:dnu}
and~\ref{eq:numax} may be used directly to infer the stellar radius,
mass, density and surface gravity (though not the age). This so-called
``direct method'' gives uncertainties that are larger than those given
by the grid-based approach because the scaling relations are not
constrained by the equations governing stellar structure and
evolution, i.e., at a given mass and radius any value of \teff\ is
possible. However, we know from stellar evolution theory that only a
narrow range of \teff\ is allowed (and moreover, the effective
temperature also depends on the chemical composition). Employing a
grid of stellar models hence gives smaller uncertainties.

A good deal of effort has recently been devoted to testing the
accuracy of the scaling relations and asteroseismically inferred
properties, using results on stars where it is possible to
independently estimate the properties to a verifiable (high) level of
accuracy (e.g., stars in binaries). For a comprehensive review, see
Chaplin \& Miglio (2013) and references therein. Examples include
comparisons with properties estimated using eclipsing binaries,
stellar parallaxes, long baseline interferometry, and members of open
clusters (e.g., Stello et al. 2008; Bedding 2011; Brogaard et
al. 2012; Miglio 2012; Miglio et al. 2012; Huber et al. 2012; Silva
Aguirre et al. 2012). Good agreement has been found for main-sequence
stars and sub-giants, with no evidence for systematic deviations found
at the level of the observational uncertainties, i.e., upper limits of
around 4\,\% in radius and 10\,\% in mass. But further tests are
needed, in particular tests of the estimated masses.

Mosser et al. (2013) have also recently advocated modifying the
observed average \dnu, for use with the scaling relations, to the
value expected in the high-frequency asymptotic limit. We test the
impact of their suggested modifications in Section~\ref{sec:res}.

Most grid-based methods determine the characteristics of stars by
finding the maximum of the likelihood function of a set of input
parameters, \{\numax, \dnu, \teff, [Fe/H]\}, calculated with respect
to a grid of stellar evolutionary models. The details of how this is
achieved varies, depending on the pipeline used. In this work we use
six different pipelines that search within eleven stellar evolutionary
grids. Some pipelines used model-calculated eigenfrequencies to
estimate the model \dnu, while a majority used the \dnu\ scaling
relation. All pipelines used the \numax\ scaling relation. The listed
stellar properties that we present later come from one of the
pipelines that used model-calculated eigenfrequencies. The pipelines
used are described further in Section~\ref{sec:pipes}.  Characteristics and
systematic biases involved with grid-based analyses have been
investigated in detail by Gai et al. (2011), Basu et al. (2012), Bazot
et al. (2012) and Gruberbauer et al. (2012).  Systematics related to
estimation of $\log g$ have recently been explored by Creevey et
al. (2013).

%%%%%%%%%%%%%%%%%%%%%%%%%%%%%%%%%%%%%%%%%%%%%%%%%%%%%%%%%%%%%%%%%%%%%%%

% Fig. 1

%%%%%%%%%%%%%%%%%%%%%%%%%%%%%%%%%%%%%%%%%%%%%%%%%%%%%%%%%%%%%%%%%%%%%%%

% Fig. 2

\begin{figure*}

\epsscale{1.0}
\plotone{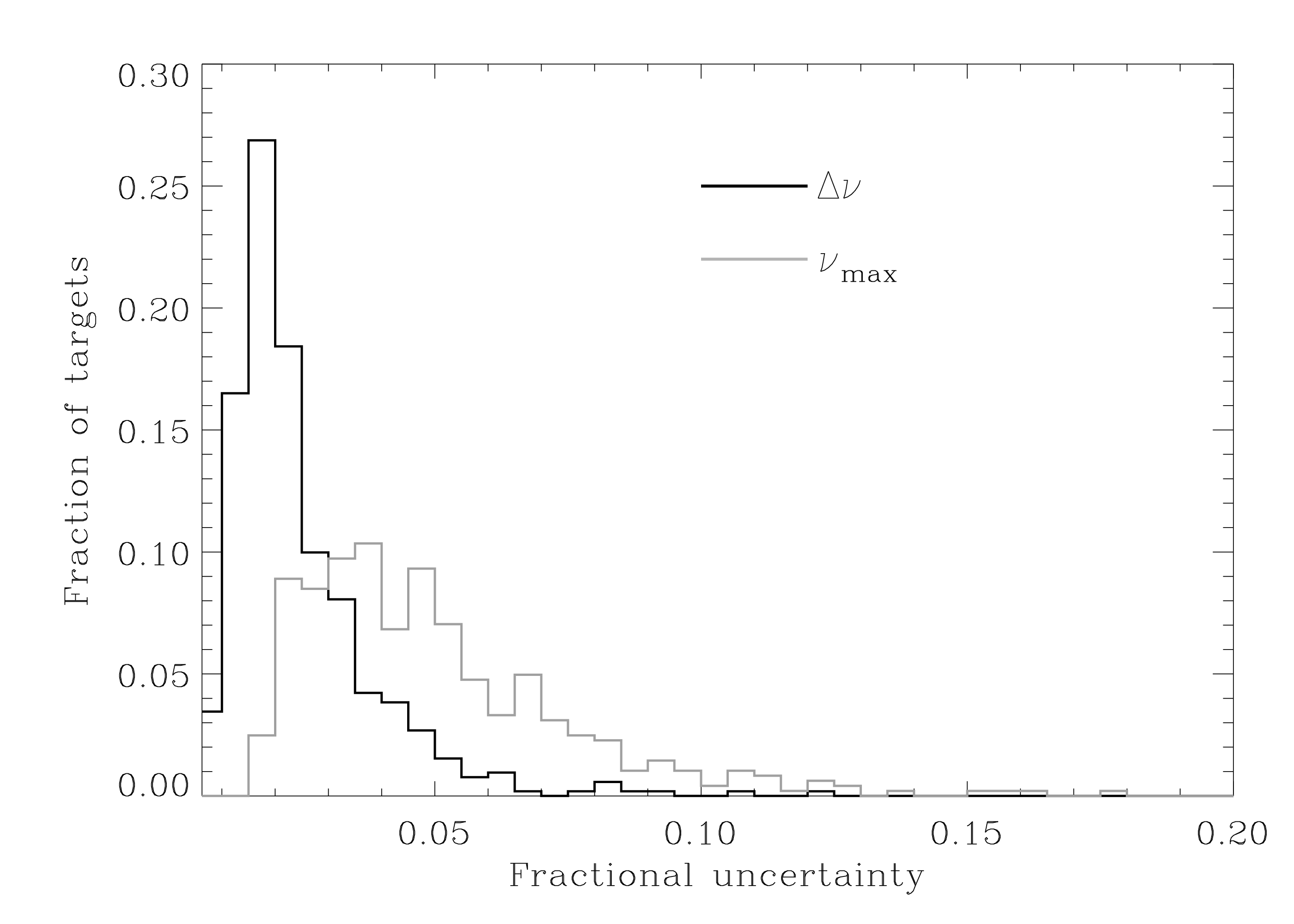}

\caption{Histograms of fractional uncertainties for the global
  asteroseismic input parameters $\Delta\nu$ and $\nu_{\rm max}$ (see
  figure legend).}

\label{fig:datahisto}
\end{figure*}

%%%%%%%%%%%%%%%%%%%%%%%%%%%%%%%%%%%%%%%%%%%%%%%%%%%%%%%%%%%%%%%%%%%%%%%

\section{Input seismic and non-seismic data}
\label{sec:inputs}

We used asteroseismic data on solar-type stars observed by
\emph{Kepler} during the first ten months of science operations. About
2000 stars, down to \emph{Kepler} apparent magnitude $K_p \simeq
12.5$, were selected as potential solar-type targets based upon
complementary data from the \emph{Kepler} Input Catalog (KIC; Brown et
al. 2011). Each target was observed continuously for one month at a
short cadence of 58.85\,sec (Gilliland et al. 2010b, Jenkins et
al. 2010). Timeseries were prepared for asteroseismic analysis in the
manner described by Garc\'ia et al. (2011). Different teams attempted
to detect, and then extract the basic properties of, the solar-like
oscillations using automated analysis pipelines developed and
extensively tested (e.g., see Bonanno et al. 2008; Huber et al. 2009;
Mosser \& Appourchaux 2009; Roxburgh 2009; Campante et al. 2010a;
Hekker et al. 2010; Mathur et al. 2010) for application to the large
ensemble of targets observed by \emph{Kepler}. The basic results of
this survey were presented in Chaplin et al. (2011) and Verner et
al. (2011).

Figure~\ref{fig:pspec} shows examples of the oscillation spectra of two
stars from the survey. The top panel is a high S/N case, where the
individual modes are easily distinguishable in the spectrum. The blue
line shows the envelope of power given by the oscillations (from
heavily smoothing the power spectrum, and here multiplied by a factor
of five to show the envelope more clearly). In this case it is trivial
to extract the average large separation, \dnu, and the frequency of
maximum oscillations power, $\nu_{\rm max}$. The bottom panel presents
a much harder, low-S/N case. Here, the S/N is much reduced because the
star is over two magnitudes fainter than the target in the top
panel. Whilst it is still possible to extract a robust estimate of
\dnu, beating of the oscillation signal with the background means
that: first, it would be much harder to extract precise estimates of
the individual mode frequencies; but also, second, and of direct
relevance to this paper, it is no longer possible to extract a
well-defined estimate of \numax. This was also the case for another 36
stars in the full cohort. Their properties, and the properties of the
star in the bottom panel of Figure~\ref{fig:pspec}, were therefore
derived with only \dnu\ used as seismic input.

We used estimates of \dnu\ and \numax\ returned by five teams. Full
details of the pipelines may be found in Verner et al. (2011),
including a discussion of the excellent level of agreement found
between the results delivered by the different codes.  The final
parameters selected for use by the grid-search pipelines were those
returned by the code described in Huber et al. (2009). Its results had
the smallest average deviation from the median values in a global
comparison made over all the stars.  Final median fractional
uncertainties were 2.1\,\% in \dnu\ and 4.6\,\% in \numax. The
distribution of fractional uncertainties in the two seismic quantities
can be seen in Figure~\ref{fig:datahisto}.

A homogeneous set of effective temperatures, $T_{\rm eff}$, was
estimated for the entire ensemble using available complementary
photometry.  One set of temperatures was derived by using an Infra-Red
Flux Method (IRFM) calibration (Casagrande et al. 2010; see also Silva
Aguirre et al. 2012).  This made use of multi-band $JHK$ photometry
from the Two Micron All Sky Survey (2MASS; Skrutskie et al. 2006),
photometry in the SDSS $griz$ bands available in the KIC, and
reddening estimates from Drimmel et al. (2003).  A second set of
temperatures were those derived by Pinsonneault et al. (2012), who
performed a recalibration of the KIC photometry in the SDSS $griz$
filters, using YREC models.  The complementary photometry that was
available to us did not allow strong constraints to be placed on the
metallicity of all the targets. When using the photometric $T_{\rm
  eff}$ in the grid searches we therefore adopted an [Fe/H]
corresponding to an average value for the field of $-0.2 \pm 0.3\,\rm
dex$ (e.g. see Silva Aguirre et al. 2011). It is worth adding that for
the IRFM, the dependence of the temperatures on [Fe/H] is rather
weak. For the SDSS calibration, a change of $\delta {\rm [Fe/H]}
\simeq 0.4$ is needed to change the temperatures at approximately the
$1\sigma$ level (see Table~3 of Pinsonneault et al. 2012).

\begin{figure*}

\epsscale{0.75}
\plotone{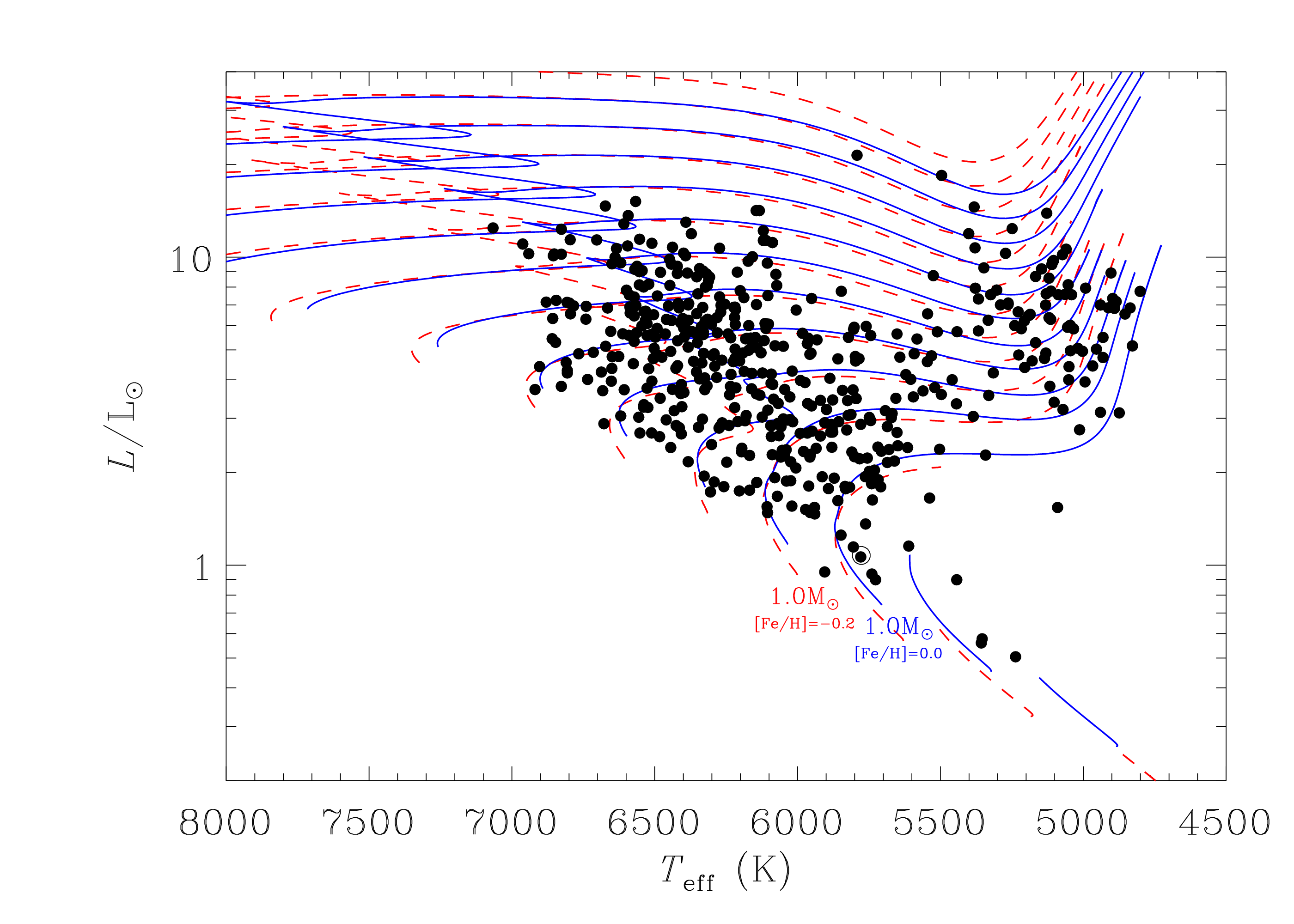}
\epsscale{0.75}
\plotone{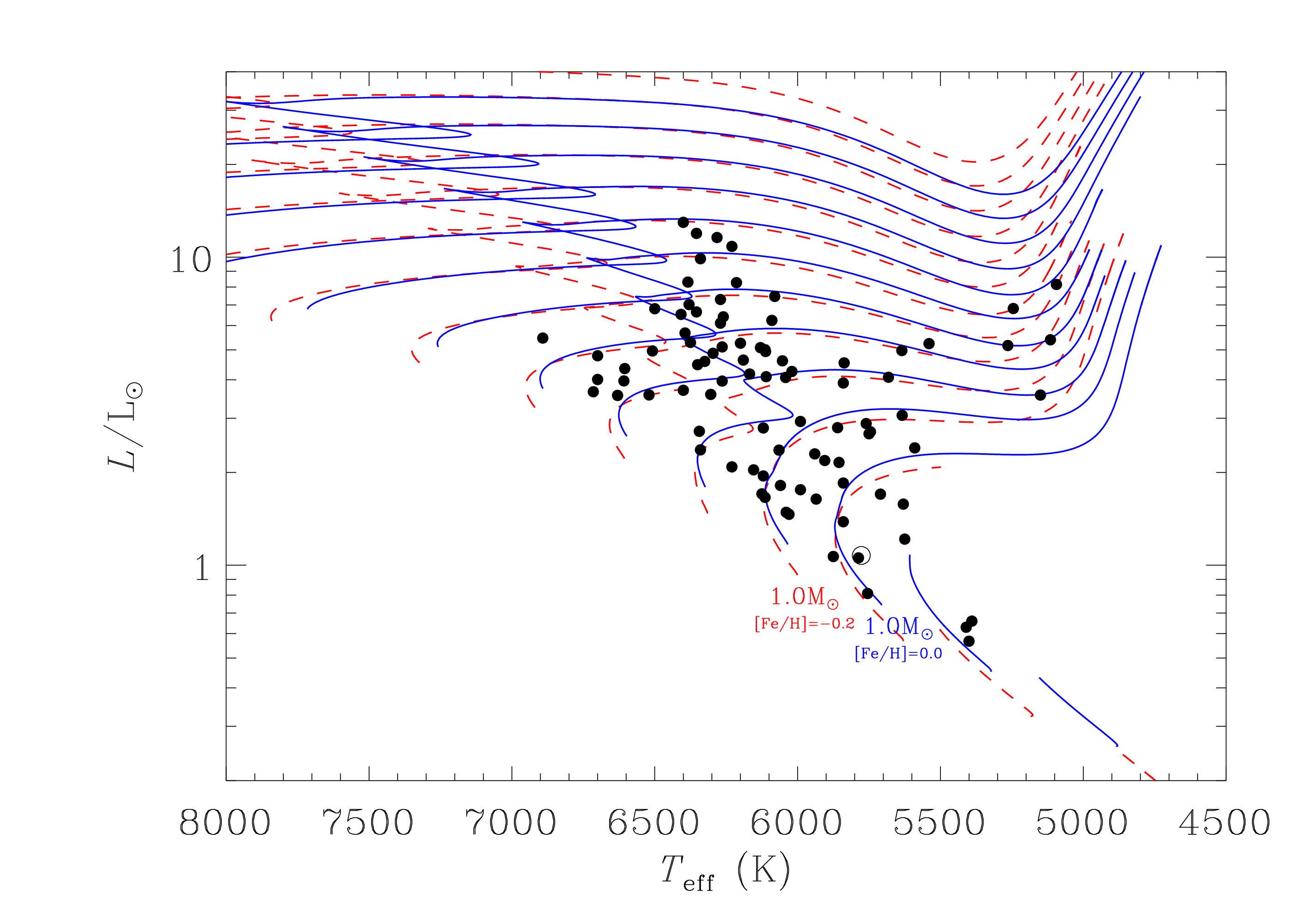}

\caption{Top panel: HR diagram of the full cohort of stars.
  Temperatures are from the IRFM set; luminosities were calculated
  using the asteroseismically estimated stellar radii presented later
  in the paper, in Table~\ref{tab:res2}. Bottom panel: HR diagram of
  the smaller cohort with spectroscopic $T_{\rm eff}$ and [Fe/H];
  luminosities were estimated using the radii presented in
  Table~\ref{tab:res3}. The evolutionary tracks in both panels were
  computed at $0.1\,\rm M_{\odot}$ intervals with the YREC
  code. Tracks in blue (solid lines) are for solar composition, those
  in red (dashed lines) for [Fe/H]$=-0.2$. The models are not ``solar
  calibrated'' models and hence the $1\,\rm M_{\odot}$ tracks need not
  pass through the exact location of the Sun.}

\label{fig:hr}
\end{figure*}

The top panel of Figure~\ref{fig:hr} shows the positions of the full
cohort of stars on an HR diagram. The temperatures are from the IRFM
set, and the luminosities were calculated using the asteroseismically
estimated stellar radii presented later in the paper.

%%%%%%%%%%%%%%%%%%%%%%%%%%%%%%%%%%%%%%%%%%%%%%%%%%%%%%%%%%%%%%%%%%%%%%%

% Fig. 3

%%%%%%%%%%%%%%%%%%%%%%%%%%%%%%%%%%%%%%%%%%%%%%%%%%%%%%%%%%%%%%%%%%%%%%%

Homogeneous sets of spectroscopic $T_{\rm eff}$ and [Fe/H] were
available on a subset of 87 stars, from the data reductions performed
by Bruntt et al. (2012) on high-resolution spectra obtained with the
ESPaDOnS spectrograph at the 3.6-m Canada-France-Hawaii Telescope
(CFHT), and with the NARVAL spectrograph mounted on the 2-m Bernard
Lyot Telescope at the Pic-du-Midi Observatory in France.  To account
for systematic differences between spectroscopic methods, we followed
the procedure suggested by Torres et al. (2012) and added in
quadrature 59\,K to all $T_{\rm eff}$ uncertainties and 0.062\,dex to
all [Fe/H] uncertainties given by Bruntt et al. (2012). The
distribution of this smaller sample of stars is plotted in HR form in
the bottom panel of Figure~\ref{fig:hr}.

All grid-modelling pipelines mentioned below were required to
determine stellar properties using \{\dnu, \numax, \teff, \feh\} as
inputs.  These seismic and non-seismic input parameters for the grid
modelling are listed in Tables~\ref{tab:data1} and~\ref{tab:data2}.
As noted above, for stars where \numax\ was uncertain, only \{\dnu,
\teff, \feh\} were used.  All pipelines were asked to use
$\Delta\nu_\odot=135.1\,\mu$Hz and $\nu_{\rm max,\odot}=3090\,\mu$Hz,
which are the reference values for the Huber et al. (2009) pipeline
derived from the analysis of VIRGO/SoHo Sun-as-a-star data (Huber et
al. 2011).  The uncertainties in $\Delta\nu_\odot$ ($0.1\,\rm \mu Hz$)
and $\nu_{\rm max,\odot}$ ($30\,\rm \mu Hz$) were accounted for by
increasing the uncertainties in \dnu\ and \numax\ by simple error
propagation.

\section{Grid-based pipelines: details}
\label{sec:pipes}

We used six different grid-based pipelines to determine stellar
properties:

 \begin{enumerate}
 \item Yale-Birmingham (YB) (Basu et al. 2010, 2012, Gai et al. 2011);

 \item Bellaterra Stellar Properties Pipeline (BeSPP) (Serenelli et
   al. 2013 in preparation);

 \item RadEx10 (Creevey et al. 2013);

 \item RADIUS (Stello et al. 2009); 

 \item SEEK (Quirion et al. 2010); and

 \item G{\"o}ttingen (GOE) (Ball et al. in preparation, details below).

 \end{enumerate}
Further details of how most of the pipelines work are available in the
literature. Here, we provide brief outlines.

The YB pipeline is based on finding the maximum likelihood of the set
of input parameter data calculated with respect to the grid of
models. For a given observational (central) input parameter set, the
first key step in the method is to generate 10,000 input parameter
sets by adding different random realizations of Gaussian noise,
commensurate with the observational uncertainties, to the actual
(central) observational input parameter set. The distribution of any
property, say radius, is then obtained from the central parameter set
and the 10,000 perturbed parameter sets, which form the distribution
function. The final estimate of the property is the median of this
distribution. The 1$\sigma$ limits from the median are adopted as
measures of the uncertainties.  The BeSPP and RadEx10 pipelines both
employ the same principles as the YB pipeline. They differ in some
minor details, and also in whether the mean or the median of the
distribution function is used as the adopted value of the property. We
have verified that this choice does not have a significant impact on
the results presented in this paper.

RADIUS follows a slightly different approach. It finds all models
whose parameters lie within 3$\sigma$ of the observations. Properties
are estimated from the properties of the most likely model, with the
1$\sigma$ uncertainties estimated as one-sixth of the maximum range of
the selected models.

SEEK compares an observed star with every model of the grid and makes
a probabilistic assessment of the stellar properties with the help of
Bayesian statistics.  Each stellar model in the grid is assigned a
posterior probability that is the product of a Gaussian likelihood for
each observable and an appropriate prior for the desired parameters.
The probabilities are normalized so that the sum over all the stellar
models is unity.  For a given property, the probabilities are then
summed in a suitable range of bins.  In effect, one constructs a
histogram of the desired property where each stellar model is weighted
by its posterior probability.  By associating the center of each bin
with its height, a probability density function (PDF) is created from
which the final values of the properties are derived.  The priors used
are flat for age, metallicity, initial helium ratio, and mixing length
parameter. The only non-flat prior is that of the initial mass
function. SEEK used \numax\ only to select models, but does not use
this parameter to obtain the final result.

The GOE pipeline is an independent implementation of the
SEEK method.  While the Bayesian method defined by the SEEK algorithm
allows for the inclusion of different types of prior information, the
G{\"o}ttingen implementation only includes priors that correct for the
non-uniform distribution of models in metallicity and age. To correct
the age distribution, each model is weighted by the time-step of that
model, so that models that are evolving more slowly are more likely.
This counteracts the fact that evolutionary codes calculate more
models in rapid phases of evolution.  Without this correction, the
results would be biased towards these rapid phases.

The YB pipeline was used with 5 different grids---the models from
the Dartmouth group (Dotter et al. 2008), those of the Padova group
(Marigo et al. 2008, Girardi et al. 2000), the models that comprise
the YY isochrones (Demarque et al.  2004), a grid of models
constructed with the Yale Stellar Evolution Code (YREC; Demarque et
al. 2008) and described by Gai et al. (2011) (we refer to this set as
YREC), and another set of models constructed with YREC (we refer to
this grid as YREC2) that has been described by Basu et al. (2012).

Although the YREC and YREC2 grids were constructed with the same code,
they have different physics. These grids were calculated using
different nuclear reaction rates, different relative heavy-element
abundances and a different helium enrichment law. Additionally, all
models in the YREC grid were calculated with the same value of the
mixing-length parameter; YREC2 on the other hand consists of five
sub-grids, each sub-grid constructed with a different value of the
mixing-length parameter. For all models in these grids, the seismic
parameter \dnu\ was calculated using the scaling relation given in
Equations~\ref{eq:dnu}.

The BeSPP pipeline was run with two grids.  The first grid comprises
 models constructed with the GARSTEC code (Weiss \& Schlattl 2008)
and the parameters of the grid are described in Silva Aguirre et
al. (2012). The \dnu\ of each model in this grid was determined using
the calculated frequencies of each model (one set of results) and also
using the scaling relation in Equation~\ref{eq:dnu} (to give a second
set of results that we call BeSPPscale). The second grid of models are
the BASTI models of Pietrinferni et al. (2004) computed specifically
for use in asteroseismic studies, and as described in Silva Aguirre et
al (2013).  In this case \dnu\ for the models was calculated using
only the scaling relation.

RadEx10, RADIUS and SEEK used models constructed with the ASTEC code
(Christensen-Dalsgaard et al. 2008).  The models used in RadEx10,
referred to as ASTEC1, are described by Creevey et
al. (2013). \dnu\ for this grid was calculated from the scaling
relation. The models for RADIUS are described in Stello et al. (2009)
and Creevey et al. (2012), and these models are henceforth referred to
as ASTEC2. While \dnu\ for this grid was calculated using the scaling
relations, it was also calculated for a subset of ASTEC2 using
individual frequencies.  The models used by SEEK are referred to as
ASTEC3. For this grid \dnu\ was calculated using eigenfrequencies.
Although all the grids were constructed with the same code, they have
different physics, such as low temperature opacities and equation of
state, and different input parameters.

GOE was run on a grid calculated with the CESTAM code
(Marques et al.~2013), which is derived from the CESAM2k code
described in Morel \& Lebreton (2008). The \dnu\ of each model was
calculated from the eigenfrequencies.

The parameters of the different grids of models are listed in
Table~\ref{tab:pipelines}. As can be seen, the grids are diverse, not
only constructed with different codes, but also with different input
physics. For example, there are grids of models with and without
diffusion and overshoot; and also grids constructed with different
model prescriptions, e.g., for overshoot and He enrichment.

\section{Comparison of results from different grid-pipeline combinations}
\label{sec:res}

Before consolidating the results to give tables of stellar properties,
we first present a comparison of the estimates returned by the various
grid-pipelines. To frame the discussion we have selected
representative plots of differences shown by certain grid-pipeline
combinations. Appendix~\ref{sec:app} shows detailed plots
(Figures~\ref{fig:diffteff1_a} through~\ref{fig:diffteff2_e}) of
differences for all the pipelines.

We begin by presenting results given by the BeSPP pipeline coupled to
the GARSTEC grid. The BeSPP/GARSTEC combination can be run in two
ways, one where \dnu\ is calculated using the eigenfrequencies, and
one where \dnu\ is calculated using the scaling relations. This allows
us to test the differences in the results caused by how \dnu\ is
calculated, independently of differences in results arising from the
analysis pipeline or the grid of stellar models used.

\begin{figure*}
\epsscale{1.0}
\plotone{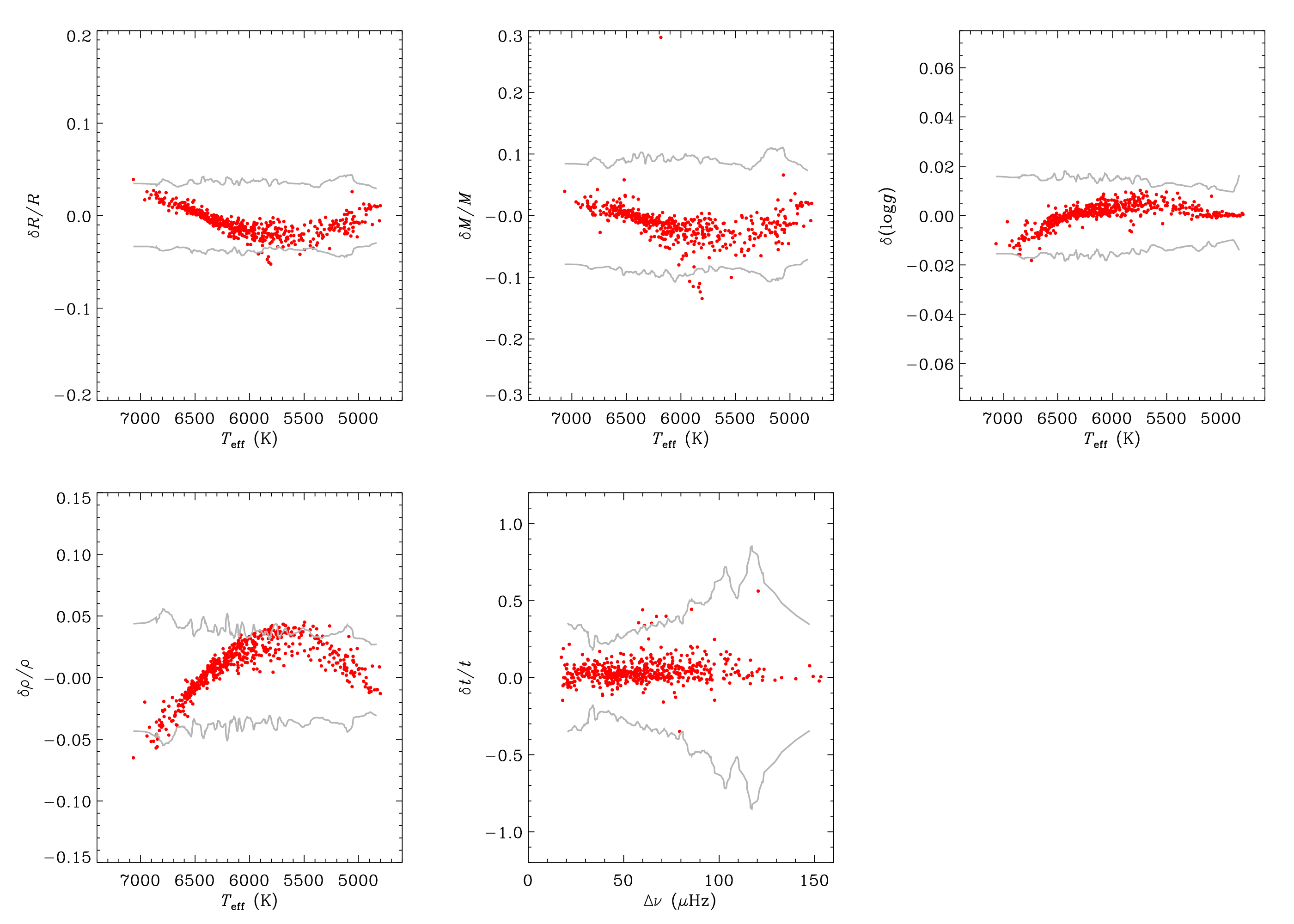}

\caption{Fractional differences in estimated stellar properties for
  analyses performed by BeSPP with the GARSTEC grid, for  the entire
  ensemble with \dnu\ and $\nu_{\rm max}$, the photometric (IRFM)
  $T_{\rm eff}$ and field [Fe/H] values used as inputs. The plots show
  differences between using model-calculated eigenfrequencies to
  estimate the \dnu\ of each model and using the \dnu\ scaling
  relation (in the sense scaling minus frequencies). Gray lines mark
  the median $1\sigma$ envelope of the grid-pipeline returned, formal
  uncertainties.  These lines are included to help judge the
  \emph{typical} precision only.}

\label{fig:diffpars1}
\end{figure*}

Figure~\ref{fig:diffpars1} shows the resulting differences in
estimated stellar properties for the entire ensemble, with \dnu\ and
$\nu_{\rm max}$, the photometric (IRFM) $T_{\rm eff}$ and field [Fe/H]
values used as inputs. The plots show differences in the sense
scaling-mode outputs minus frequency-mode outputs, and are fractional
differences in $R$, $M$, $\rho$ and age $t$; and absolute differences
in $\log\,g$. Age differences have been plotted against \dnu\ to
delineate approximately the evolutionary state (since \dnu\ gives a
first-order discrimination of main-sequence, sub-giant and
low-luminosity red-giant targets). The gray lines mark envelopes
corresponding to the median of the formal $1\sigma$ uncertainties
returned by all grid-pipelines. Medians were calculated in 10-target
batches sorted on the independent variable used for the plots ($T_{\rm
  eff}$ for $R$, $M$, $\log\,g$ and $\rho$; and \dnu\ for $t$). These
lines are included to help judge the \emph{typical} precision only;
the uncertainties in the results of any particular star may be
slightly different.

We see clearly the impact of adopting the scaling relation to compute
\dnu\, instead of using model-calculated eigenfrequencies. The
``boomerang'' shaped trends arise directly from the similar-shaped
differences shown between \dnu\ calculated using the scaling relation
and the individual eigenfrequencies, as discussed earlier in
Section~\ref{sec:prin}. The impact is strongest in the estimated
densities. The boomerang shape is absent from the age differences,
although there is a small positive bias arising from the negative
differences displayed in the masses. It is worth noting that, at the
level of precision of these data, the boomerang-shaped differences lie
largely within the median $1\sigma$ uncertainty envelopes.

\begin{figure*}
\epsscale{1.0}
\plotone{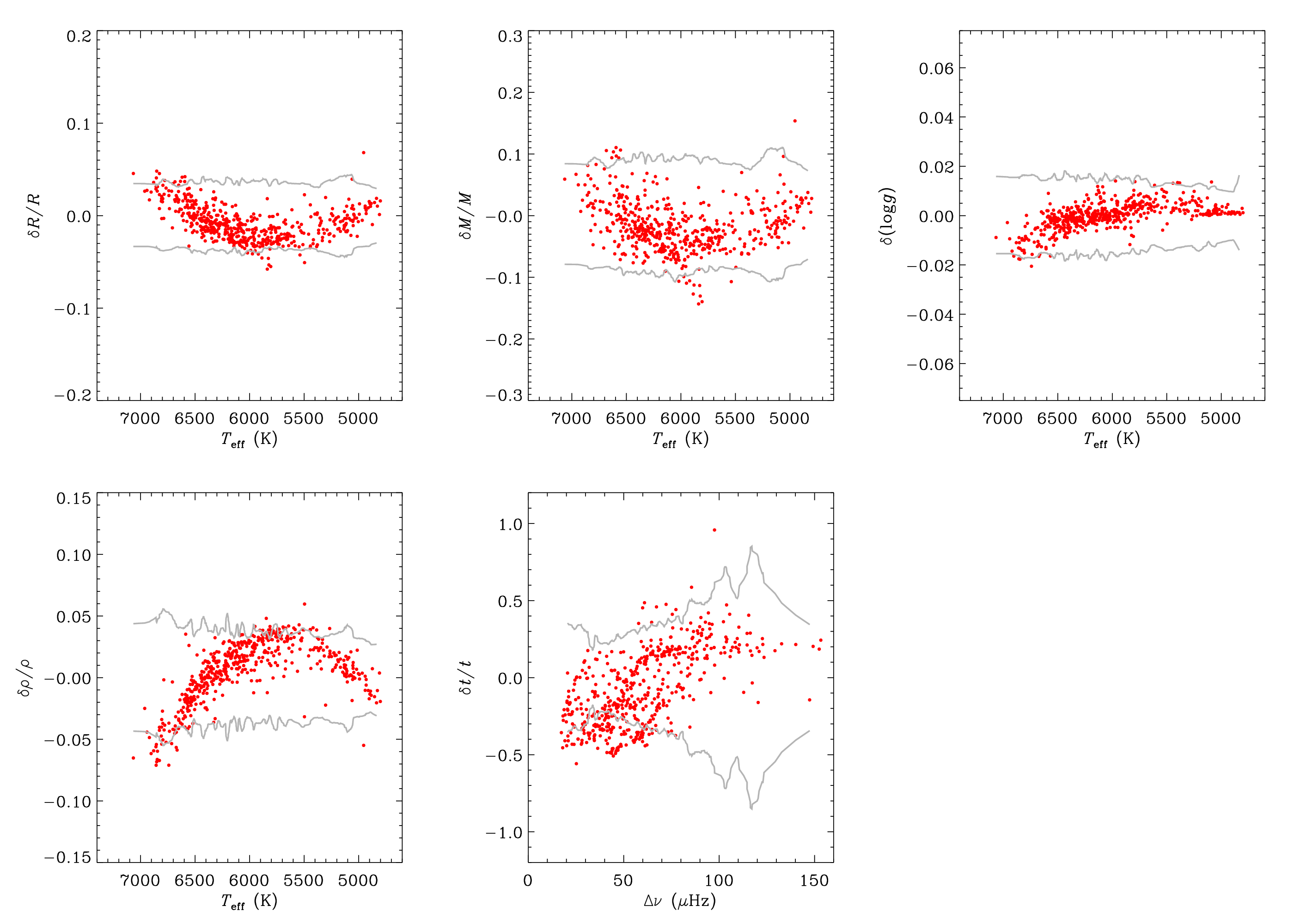}

\caption{Fractional differences in estimated stellar properties
  between BeSPP/BASTI (scaling mode) and BeSPP/GARSTEC (run in
  frequency mode). Results shown for the entire ensemble, with
  \dnu\ and $\nu_{\rm max}$, the photometric (IRFM) $T_{\rm eff}$ and
  field [Fe/H] values used as inputs). Plot style as per
  Figure~\ref{fig:diffpars1}.}

\label{fig:diffpars2}
\end{figure*}

\begin{figure*}
\epsscale{1.0}
\plotone{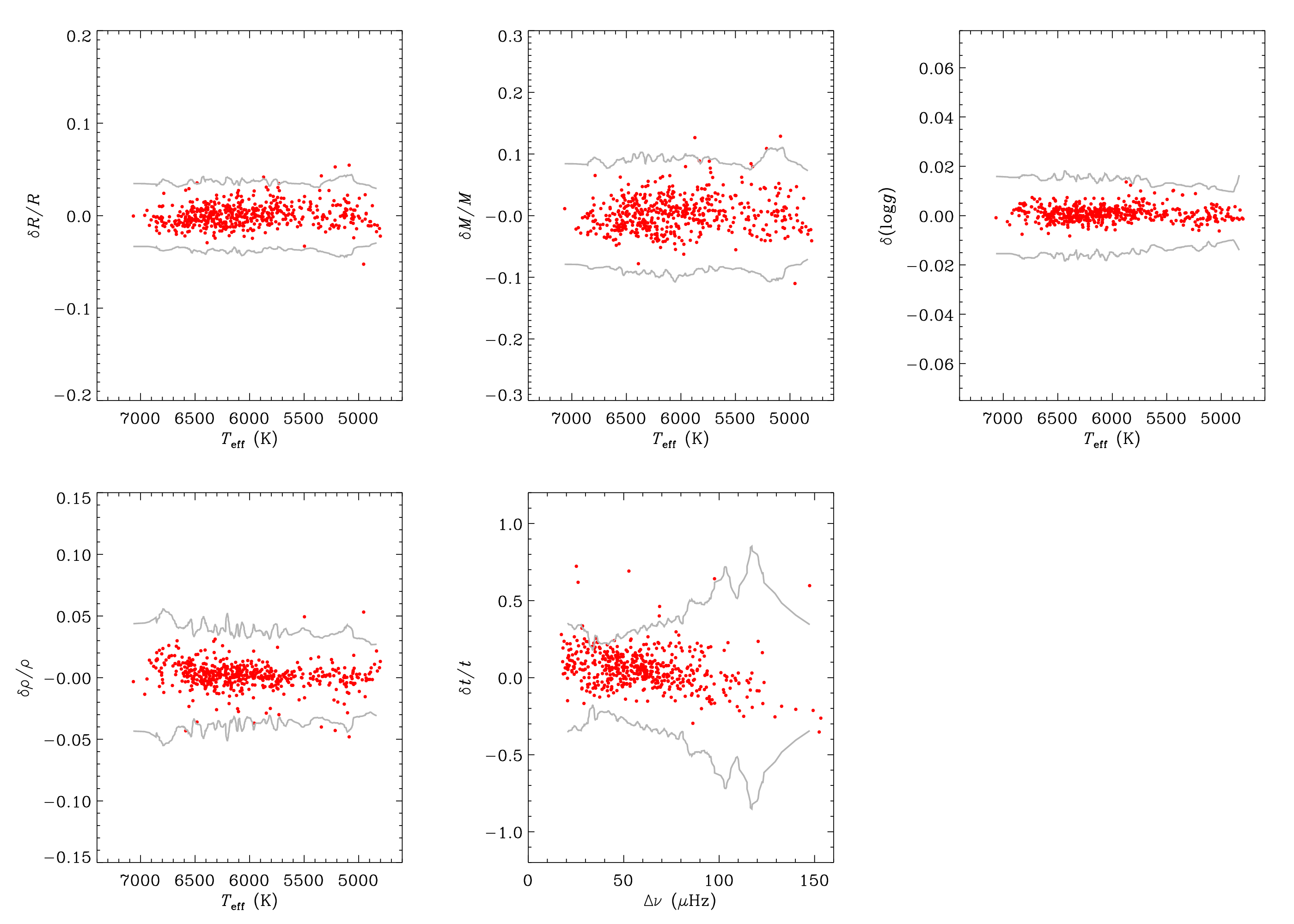}

\caption{Fractional differences in estimated stellar properties
  returned by the BeSPP pipeline (run in scaling mode) and the YB
  pipeline, but with both coupled to the same YY grid. Results shown
  for the entire ensemble, with \dnu\ and $\nu_{\rm max}$, the
  photometric (IRFM) $T_{\rm eff}$ and field [Fe/H] values used as
  inputs). Plot style as per Figure~\ref{fig:diffpars1}.}

\label{fig:diffpars3}
\end{figure*}

In addition to the impact of the scaling relation, we also expect
differences in results due to the choice of grid of stellar
evolutionary models, and the actual pipelines themselves, i.e., due to
differences in methodology and procedure. First,
Figure~\ref{fig:diffpars2} shows a representative example of changing
grids. Here, we plot differences between BeSPP/BASTI (scaling-mode)
and BeSPP/GARSTEC (frequency-mode). The boomerang-shaped trends from
Figure~\ref{fig:diffpars1} are still present, but there is now
increased scatter due to differences between the grids, i.e., model
dependencies in the results. This increased level of scatter is also
present in differences between other grid-pipeline combinations (see
plots in Appendix~\ref{sec:app}). Next, we isolate the impact of
scatter due to different fitting methodologies by coupling different
pipelines to the same grid of models. Figure~\ref{fig:diffpars3} shows
a representative example, where we coupled the BeSPP pipeline to the
YY grid in order to calculate the plotted differences, between
BeSPP/YY and YB/YY. Although the differences lie fairly comfortably
within the formal uncertainty envelopes, they are not entirely
negligible (note how the ages show a small systematic bias in \dnu).
Tests of the grid-based and pipeline-based errors, using other
grid-pipeline combinations at our disposal, indicate that differences
given by the choice of grid are typically more important than those
given by the pipeline code.  Our consolidation of results to give
final uncertainties on the stellar properties includes the effects of
both error contributions (see Section~\ref{sec:tables}). Again, we
stress that on the whole these combined differences lie within the
median $1\sigma$ uncertainty envelopes.

It is not surprising that of all the properties, the ages for the full
cohort are the most scattered and poorly constrained. They also have
the largest formal fractional uncertainties. Neither \dnu\ nor
\numax\ contain any explicit dependence on age; and for the full
cohort we lack strong constraints on the metallicities, which
determine how fast a star evolves and also the effective temperature
for a given luminosity. We see clear evidence of the uncertainties
dropping in more evolved stars, i.e., the lower \dnu\ stars that have
evolved off the main-sequence (see the median formal uncertainty
envelopes in the plots). This is consistent with the results found by
Gai et al. (2011). The reason for the lower uncertainty is easy to
understand. The subgiant phase is rapid and \dnu\ and $\nu_{\rm max}$,
as well as $T_{\rm eff}$, change much more rapidly than on the main
sequence, thereby giving a better determination of the age for stars
in this phase of evolution.  Use of individual frequencies, or small
frequency separations involving dipole ($l=1$) and quadrupole ($l=2$)
modes, would lead to a considerable tightening of the age
uncertainties of the main-sequence stars (see e.g.,
Christensen-Dalsgaard 1993, Cunha et al. 2007, Chaplin \& Miglio 2013;
and references therein).

The results for the subset of stars having complementary spectroscopic
inputs are also encouraging (see Appendix~\ref{sec:app} for detailed
plots). Superior constraints on [Fe/H] and \teff\ translate, as
expected, to higher precision in the estimated properties. There is
also a slightly higher fraction of differences lying outside the
median $1\sigma$ error envelopes.  Nevertheless, consistency between
the pipelines remains good.  The scatter between pipelines is again,
not surprisingly, largest for the age estimates, where the reduction
in the input errors has brought the model-dependencies of stellar age
estimates to the fore.

%%%%%%%%%%%%%%%%%%%%%%%%%%%%%%%%%%%%%%%%%%%%%%%%%%%%%%%%%%%%%%%%%%%%%%%

% Fig. 4

%%%%%%%%%%%%%%%%%%%%%%%%%%%%%%%%%%%%%%%%%%%%%%%%%%%%%%%%%%%%%%%%%%%%%%%

% Fig. 5

%%%%%%%%%%%%%%%%%%%%%%%%%%%%%%%%%%%%%%%%%%%%%%%%%%%%%%%%%%%%%%%%%%%%%%%

%%%%%%%%%%%%%%%%%%%%%%%%%%%%%%%%%%%%%%%%%%%%%%%%%%%%%%%%%%%%%%%%%%%%%%%

% Fig. 6

%%%%%%%%%%%%%%%%%%%%%%%%%%%%%%%%%%%%%%%%%%%%%%%%%%%%%%%%%%%%%%%%%%%%%%%

We also checked the impact on the results of omitting \numax\ from the
input data (i.e., using \dnu\ as the only seismic input), both for the
same pipeline coupled to different grids, and different pipelines
coupled to the same grid. We find that changes to the estimated
properties---i.e., differences between properties given by \dnu\ and
\numax, and \dnu\ alone---are less than $1\sigma$ for most of the
stars. These differences are found to be largely due to differences in
the grids, not the pipelines.

Another source that can contribute to differences in the estimated
properties is the input set of effective temperatures, $T_{\rm
  eff}$. Note that in Section~\ref{sec:tables} we provide estimated
stellar properties for each set of input $T_{\rm eff}$.

\begin{figure*}
\epsscale{1.0}
\plotone{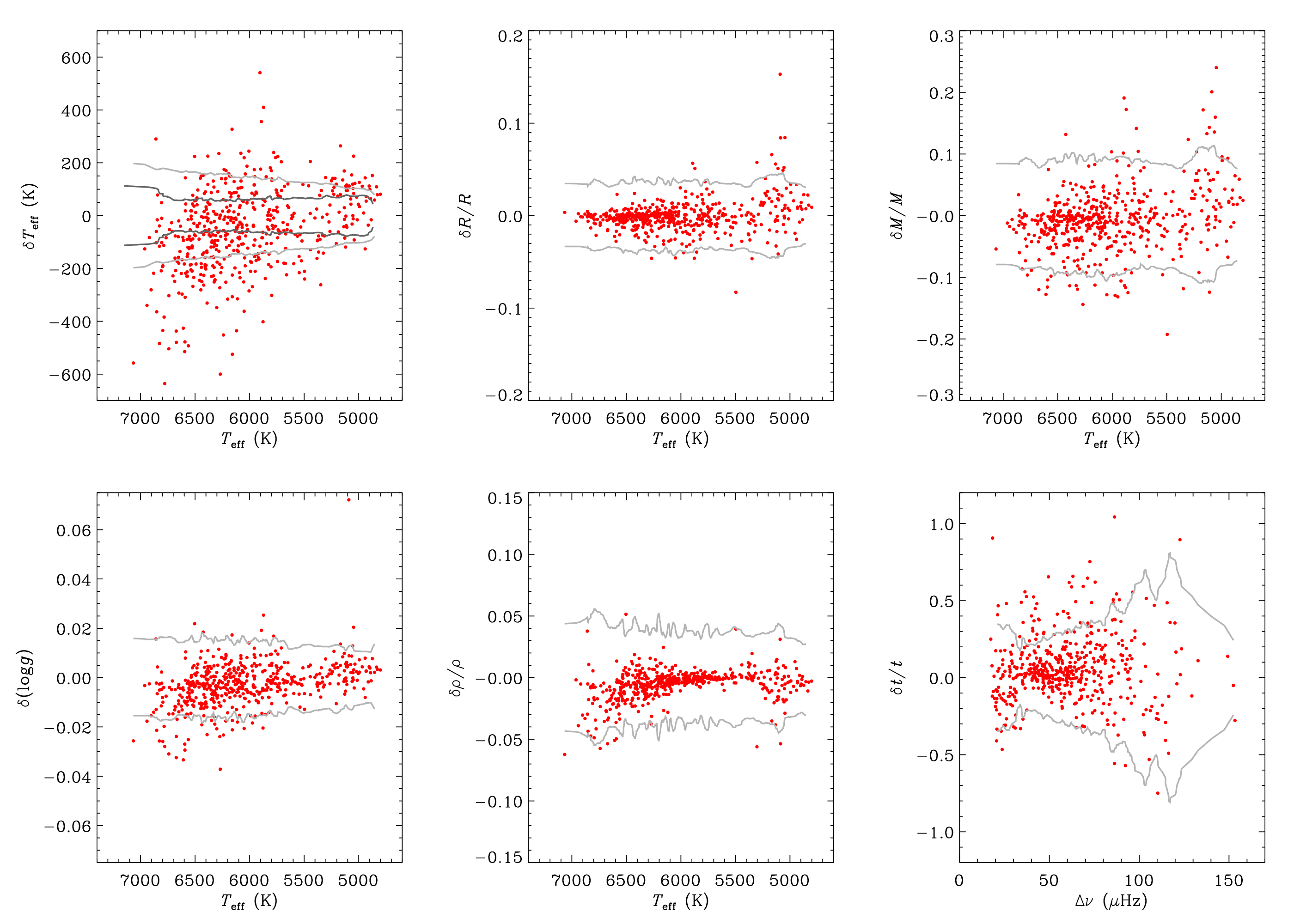}

\caption{Fractional differences in estimated properties returned by
  the BeSPP pipeline run with the GARSTEC grid (run using
  model-calculated eigenfrequencies to estimate the \dnu\ of each
  model in the grid) for analyses performed on the entire ensemble
  with different $T_{\rm eff}$ as inputs. Differences are plotted in
  the sense: results with SDSS-calibrated $T_{\rm eff}$ minus results
  with IRFM-calculated $T_{\rm eff}$. Gray lines mark the median
  $1\sigma$ envelope (over all pipelines) of the returned, formal
  uncertainties. The top left-hand panel shows the absolute
  temperature differences (same sense), with the gray lines following
  the median $1\sigma$ envelope of the IRFM uncertainties, and the
  black lines the envelope of the SDSS-calibrated uncertainties.}

\label{fig:diffteffin}
\end{figure*}

Figure~\ref{fig:diffteffin} shows the impact on the full-ensemble
results of switching from one set of input photometric $T_{\rm eff}$
to the other. The top left-hand panel shows SDSS-calibrated $T_{\rm
  eff}$ minus the IRFM-calculated $T_{\rm eff}$. The gray lines follow
the median $1\sigma$ envelope of the IRFM uncertainties, and the black
lines the envelope of the SDSS-calibrated uncertainties. The other
panels plot the fractional differences in estimated properties
returned by the BeSPP pipeline, with differences plotted in the sense
SDSS minus IRFM. As in the previous figures, gray lines mark the
median $1\sigma$ envelopes (over all pipelines) of the returned,
formal uncertainties from the IRFM results. Differences between the
two sets of temperatures may have some of their origin in differences
in the adopted reddening: Pinsonneault et al. used the reddening
information in the KIC to derive the SDSS temperatures, whilst we used
the reddening maps of Drimmel et al. (2003) to derive the IRFM
temperatures.

We may use the simple scaling relations to help us understand the
trends revealed in Figure~\ref{fig:diffteffin}. The relations imply
that $M \propto T_{\rm eff}^{1.5}$, $R \propto T_{\rm eff}^{0.5}$ and
hence $g \propto T_{\rm eff}^{0.5}$ (all other things being
equal). The trend in the $T_{\rm eff}$ is such that the SDSS
temperatures are on average slightly lower than the IRFM temperatures,
most notably at high $T_{\rm eff}$, whilst the differences are
slightly reversed at lower $T_{\rm eff}$. This trend seems to be
reflected in the plotted property differences (most notably in the
masses), which show a small negative average bias. The differences
again fall largely within the $1\sigma$ uncertainty envelopes.

\begin{figure*}
\epsscale{1.0}
\plotone{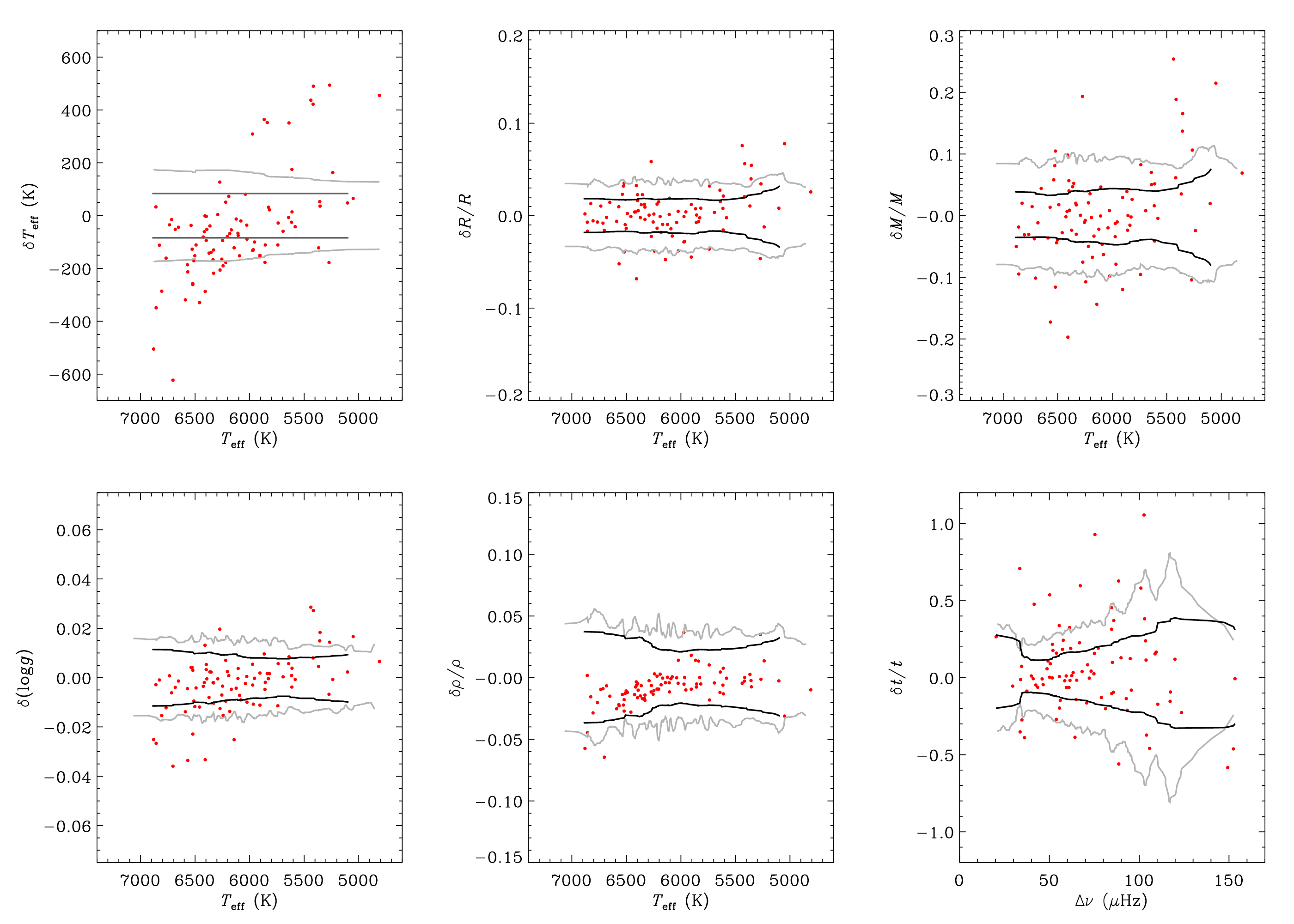}

\caption{Fractional differences in estimated properties returned by
  the BeSPP pipeline run with the GARSTEC grid, for analyses performed
  on the subset of stars with spectroscopic $T_{\rm eff}$ and [Fe/H]
  available. Differences are plotted in the sense: results with
  spectroscopic $T_{\rm eff}$ and [Fe/H] minus results with IRFM
  $T_{\rm eff}$ and field-average [Fe/H].  Gray lines mark the median
  $1\sigma$ envelope (over all pipelines) of the returned, formal
  uncertainties of the IRFM-based results, the black lines the median
  envelopes given by the spectroscopic-based results. The top
  left-hand panel shows the absolute temperature differences (same
  sense), with the gray lines following the median $1\sigma$ envelope
  of the IRFM uncertainties, and the black lines the envelope of the
  Bruntt et al. (2012) spectroscopic $T_{\rm eff}$ uncertainties.}

\label{fig:diffspecin}
\end{figure*}

Figure~\ref{fig:diffspecin} shows the impact of switching the input data
from the photometric to the spectroscopic $T_{\rm eff}$ and
[Fe/H]. The top left-hand panel shows the spectroscopic $T_{\rm eff}$
minus the IRFM-calculated $T_{\rm eff}$. The gray lines follow the
median $1\sigma$ envelope of the IRFM uncertainties, and the black
lines mark the envelope of the spectroscopic uncertainties. The other
panels plot the fractional property differences returned by BeSPP
(sense spectroscopic minus IRFM). Gray lines mark the median
uncertainty envelopes from the IRFM-based results, the black lines the
median envelopes given by the spectroscopic based-results. The trend
in the temperature differences is quite similar to that shown in
Figure~\ref{fig:diffteffin}, and the plots of the property differences
again show a small negative bias, as expected. Not surprisingly, most
differences lie well within the IRFM-based uncertainties (which we
recall used the poorly constrained field-average [Fe/H]).

%%%%%%%%%%%%%%%%%%%%%%%%%%%%%%%%%%%%%%%%%%%%%%%%%%%%%%%%%%%%%%%%%%%%%%%

% Fig. 7

%%%%%%%%%%%%%%%%%%%%%%%%%%%%%%%%%%%%%%%%%%%%%%%%%%%%%%%%%%%%%%%%%%%%%%%

% Fig. 8

%%%%%%%%%%%%%%%%%%%%%%%%%%%%%%%%%%%%%%%%%%%%%%%%%%%%%%%%%%%%%%%%%%%%%%%

Finally in this section, we note that Mosser et al. (2013) have
recently discussed modifying the observed average \dnu, for use with
the scaling relations, to the value expected in the high-frequency
asymptotic limit. The solar reference \dnu\ must also be modified.  We
have tested the impact on our results of applying this procedure, and
find that it has a negligible impact on the properties estimated by
those pipelines that use the scaling-relation-computed \dnu\ for the
stellar models.  The changes in mass and radius are typically less
than 1\,\%.  The reason for the insignificant changes is as follows.
The average correction for solar type stars is (fractionally speaking)
quite small, and is basically offset by the similar fractional
increase in the solar reference \dnu. While the fractional
modification to \dnu\ is not the same for all stars, it is a fairly
weak function of \numax, and hence luminosity $L$ (the correction is
proportional approximately to \numax$^{-0.21}$, and $\nu_{\rm max}
\propto M T_{\rm eff}^{3.5}/L^{-1}$). Thus, the modification does not
significantly affect the results (see also Hekker et al. 2013).

\section{Tables of asteroseismically inferred stellar properties}
\label{sec:tables}

\begin{figure*}

\epsscale{0.9}
\plotone{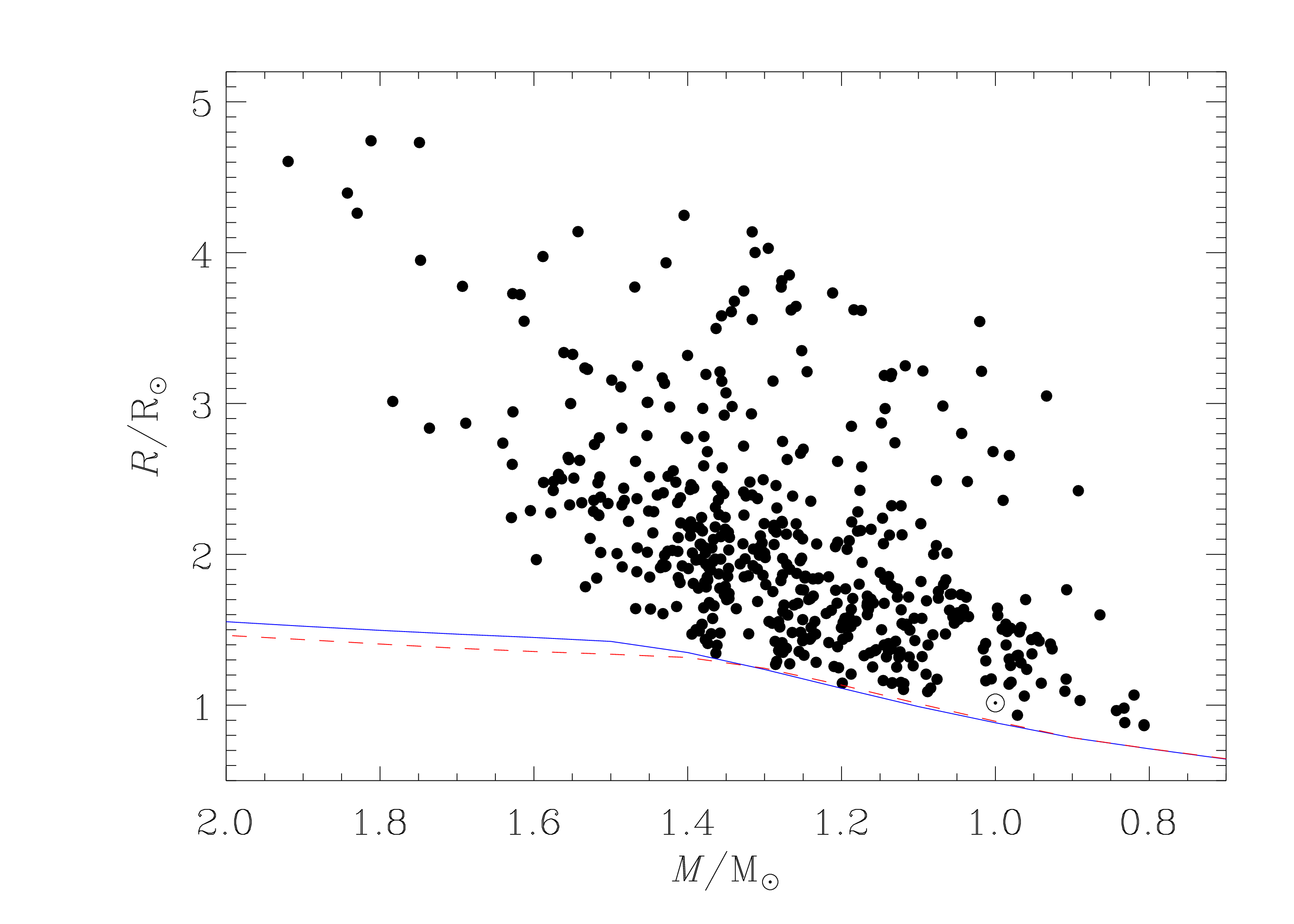}
\epsscale{0.9}
\plotone{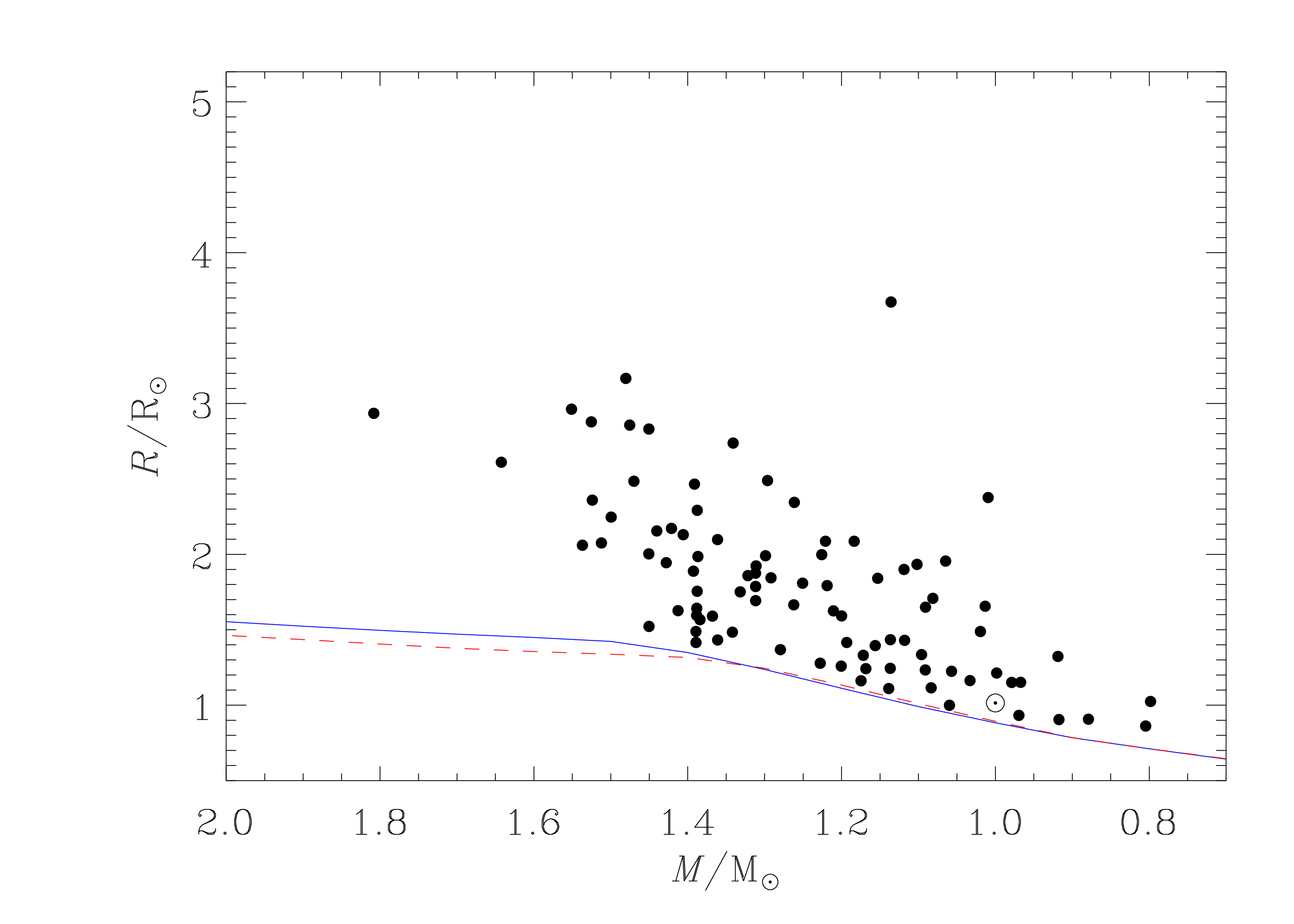}

\caption{Estimated masses and radii for the full cohort with IRFM
  inputs (top panel) and the smaller cohort with spectroscopic inputs
  (bottom panel). The solid and dashed lines mark the ZAMS for
  [Fe/H]$=0$ and $-0.2$, respectively (computed using the YREC code).}

\label{fig:mr}
\end{figure*}

Tables~\ref{tab:res1},~\ref{tab:res2} and~\ref{tab:res3} list
estimated stellar properties from our analyses. Each table gives
properties for a different set of $T_{\rm eff}$ and [Fe/H] inputs
(SDSS-calibrated $T_{\rm eff}$ and field-average [Fe/H] values; IRFM
$T_{\rm eff}$ and field-average [Fe/H] values; and Bruntt et
al. spectroscopic values, respectively). Figure~\ref{fig:mr} plots
results given by the IRFM inputs (top panel) and spectroscopic inputs
(bottom panel) in the mass-radius plane, with the location of the ZAMS
also plotted for solar and sub-solar [Fe/H] (see caption).

These final properties come from coupling BeSPP to the GARSTEC grid,
run in the mode where theoretical oscillation frequencies of each
model were used to compute average model \dnu. This meant that the
final properties are not subject to the small bias in the
\dnu\ scaling relation (manifested as the boomerang-shaped plot
differences). Details of the adopted input physics for GARSTEC are
given in Table~\ref{tab:pipelines}.

The uncertainties for the properties of each star listed in
Tables~\ref{tab:res1},~\ref{tab:res2} and~\ref{tab:res3} were given by
adding (in quadrature) the uncertainty returned by the BeSPP pipeline
to the standard deviation (scatter) of the star's property over all
grid-pipeline combinations. By including this scatter in the error
budget we capture explicitly the uncertainties arising from
differences between the commonly used grids of models we have adopted,
and scatter due to different methodologies from the different pipeline
codes.

%%%%%%%%%%%%%%%%%%%%%%%%%%%%%%%%%%%%%%%%%%%%%%%%%%%%%%%%%%%%%%%%%%%%%%%

% Fig. 9

%%%%%%%%%%%%%%%%%%%%%%%%%%%%%%%%%%%%%%%%%%%%%%%%%%%%%%%%%%%%%%%%%%%%%%%

% Fig. 10

\begin{figure*}

\epsscale{1.0}
\plotone{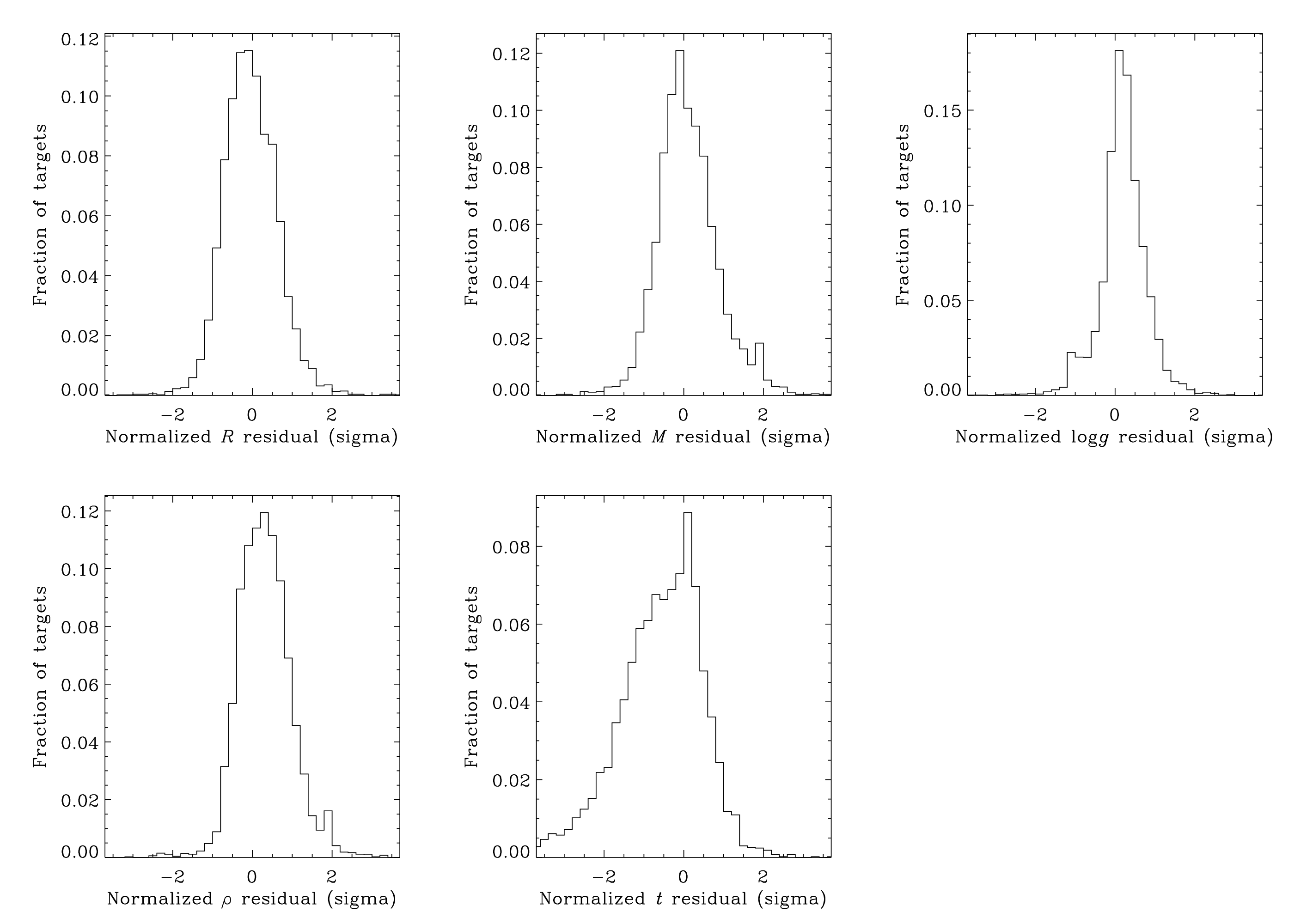}

\caption{Histograms, for each property, of uncertainty-normalized
  residuals over all grid-pipeline combinations (omitting
  BeSSP/GARSTEC) and all stars in the IRFM cohort (see
  text). Residuals calculated with respect to the BeSSP/GARSTEC
  results run in frequency mode.}

\label{fig:superdis}
\end{figure*}

%%%%%%%%%%%%%%%%%%%%%%%%%%%%%%%%%%%%%%%%%%%%%%%%%%%%%%%%%%%%%%%%%%%%%%%

Distributions of the scatter between pipelines are shown in
Figure~\ref{fig:superdis}. The histograms, which show results for the
IRFM results on the full cohort, were produced as follows. Omitting
results from BeSSP/GARSTEC, for each star we computed residuals for
each grid-pipeline with respect to the BeSSP/GARSTEC result (frequency
mode), normalizing each residual by the median property uncertainty
given by the pipelines for that star. We then accumulated residuals
for all stars in the cohort, and binned the residuals to give the
plotted ``super distributions''.

The most striking aspect of the histograms is their Gaussian-like
appearance (but see comments below on the slightly skewed age
distribution). This indicates that when a wide selection of grids is
used, differences in the input ingredients and physics give, to first
order, normally distributed-like scatter. This lends weight to our
approach of including this scatter contribution in our final error
budget by using the standard deviation. Note that residuals for the
gravities are the most peaked, possibly suggesting that the formal
uncertainties in $\log g$ are slight overestimates. The residuals in
$\rho$ are slightly offset, and this probably has a contribution from
the small offset given by the \dnu\ scaling relation (recall that most
of the pipelines rely on use of the scaling relation, which the
BeSSP/GARSTEC reference here does not). The residuals for $M$ are also
slightly off-center, though at a level much smaller than the
uncertainty in the results.

\begin{figure*}

\epsscale{1.0}
\plotone{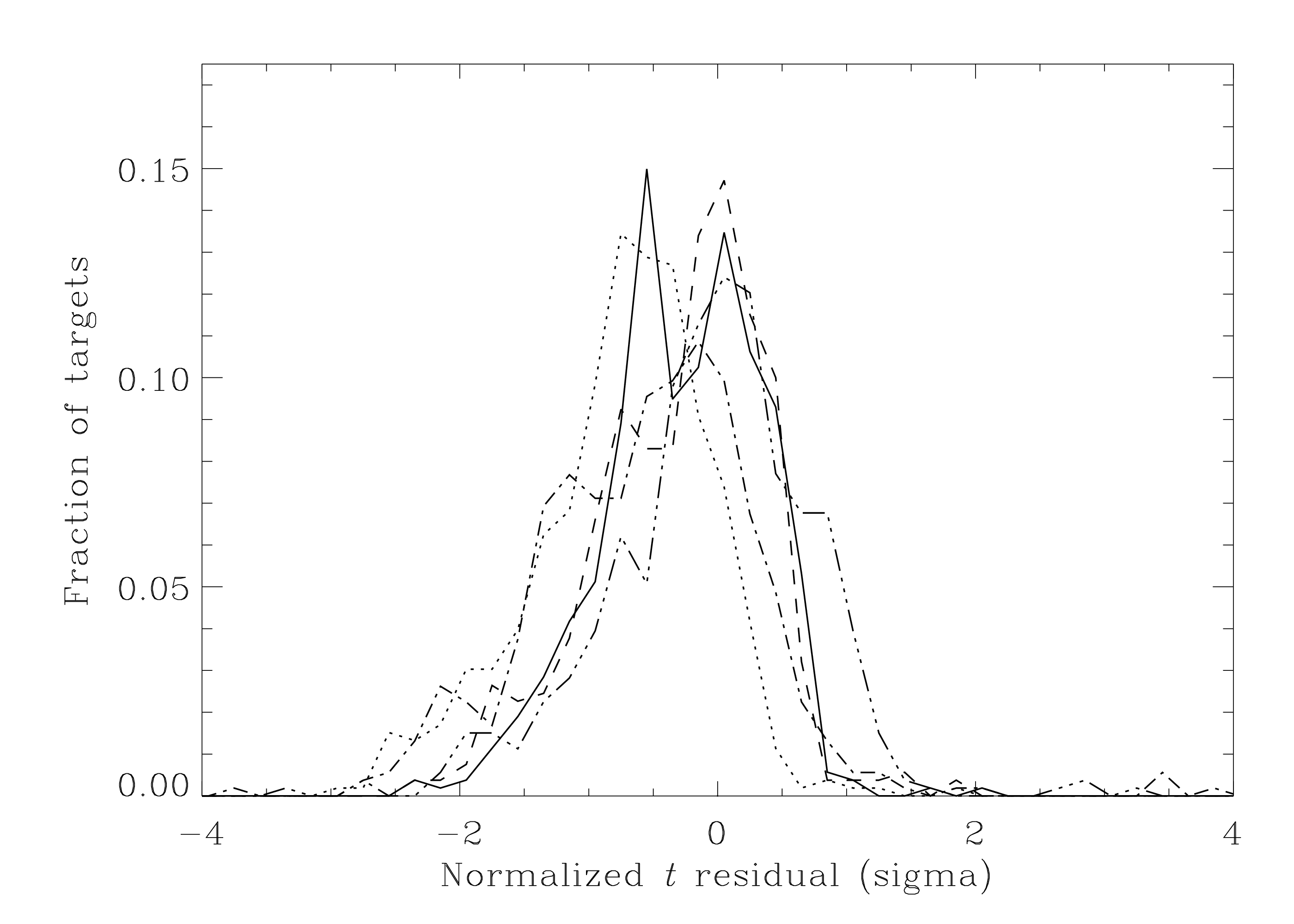}

\caption{Histogram for estimated ages, in normalized residual form as
  per the age histogram in Figure~\ref{fig:superdis}, but showing
  results for each of coupled to the YB pipeline. Residuals again
  calculated with respect to the BeSSP/GARSTEC results run in
  frequency mode. Note that for clarity we have plotted each of the
  histograms by joining the midpoints of the bins.}

\label{fig:ybt}
\end{figure*}

The distribution of age-residuals is of course the most interesting.
Normalizing the residuals by the uncertainties means that for the full
cohort the errors in the results caused by the lack of metallicity
data cannot be the cause of the skewness of the distribution (recall
that here we plot the IRFM results). Plots of results from the smaller
cohort with spectrosopic data show a similar-shaped distribution.  The
non-Gaussian nature of the distribution is mainly a result of the fact
that, of all the properties, model dependences are most marked in the
ages, i.e., differences in the physics of the grids result in
different ages for the same inputs. These dependencies were explored
in detail by Gai et al. (2011), and we see similar systematic effects
here. Figure~\ref{fig:ybt} helps to illustrate the grid-based
systematics associated with the ages. It shows histograms of the
normalized residuals in $t$---as per the normalized residuals in
Figure~\ref{fig:superdis}---but for individual grids, and only those
coupled to the same (YB) pipeline. Note that for clarity we have
plotted each of the histograms by joining the midpoints of the
bins. The offsets between histograms indicate that the systematics are
no larger than the median formal uncertainties returned by the
pipelines.

The more pronounced negative tail in the age histogram in
Figure~\ref{fig:superdis} basically tells us that ages determined by
GARSTEC are in general slightly higher than the average.

%%%%%%%%%%%%%%%%%%%%%%%%%%%%%%%%%%%%%%%%%%%%%%%%%%%%%%%%%%%%%%%%%%%%%%%

% Fig. 11

%%%%%%%%%%%%%%%%%%%%%%%%%%%%%%%%%%%%%%%%%%%%%%%%%%%%%%%%%%%%%%%%%%%%%%%

For the full cohort, the median standard deviations (scatter between
grid-pipelines)---which provide a measure of the combined effect of
the grid-based and pipeline-based errors---are approximately 4.5\,\%
in mass, 1.7\,\% in radius, 0.006\,dex in $\log\,g$, 1.3\,\% in
density, and 16\,\% in age (similar for both the IRFM and SDSS
results); for the smaller cohort with spectroscopic data they are
around 3.7\,\%, 1.3\,\%, 0.005\,dex, 1.2\,\%, and 12\,\%,
respectively.

The above measures of scatter are combined in quadrature with the
individual formal uncertainties to yield the final uncertainties on
the estimated properties.  Owing to the much tighter constraints on
[Fe/H], the final uncertainties for the sample with spectroscopic data
are, as expected, smaller than those on the full cohort. Median final
uncertainties on the spectroscopic sample are $\approx 5.4$\,\% in
mass, 2.2\,\% in radius, 0.010\,dex in $\log\,g$, 2.8\,\% in density,
and 25\% in age; results on \emph{common} stars from the full cohort
yield final median uncertainties of around 9.4\,\%, 3.5\,\%,
0.015\,dex, 3.3\,\%, and 32\,\%, respectively for the IRFM results,
and slightly smaller values for the SDSS results (owing to the
slightly smaller fractional $T_{\rm eff}$ uncertainties for the SDSS
data). When all stars in the full cohort are taken, median final
uncertainties for the IRFM results are $\approx 10.8$\,\% in mass,
4.4\,\% in radius, 0.017\,dex in $\log\,g$, 4.3\,\% in density, and
34\% in age (again, slightly smaller for the SDSS results).

%%%%%%%%%%%%%%%%%%%%%%%%%%%%%%%%%%%%%%%%%%%%%%%%%%%%%%%%%%%%%%%%%%%%%%%

% Fig. 12

\begin{figure*}

\epsscale{1.0}
\plotone{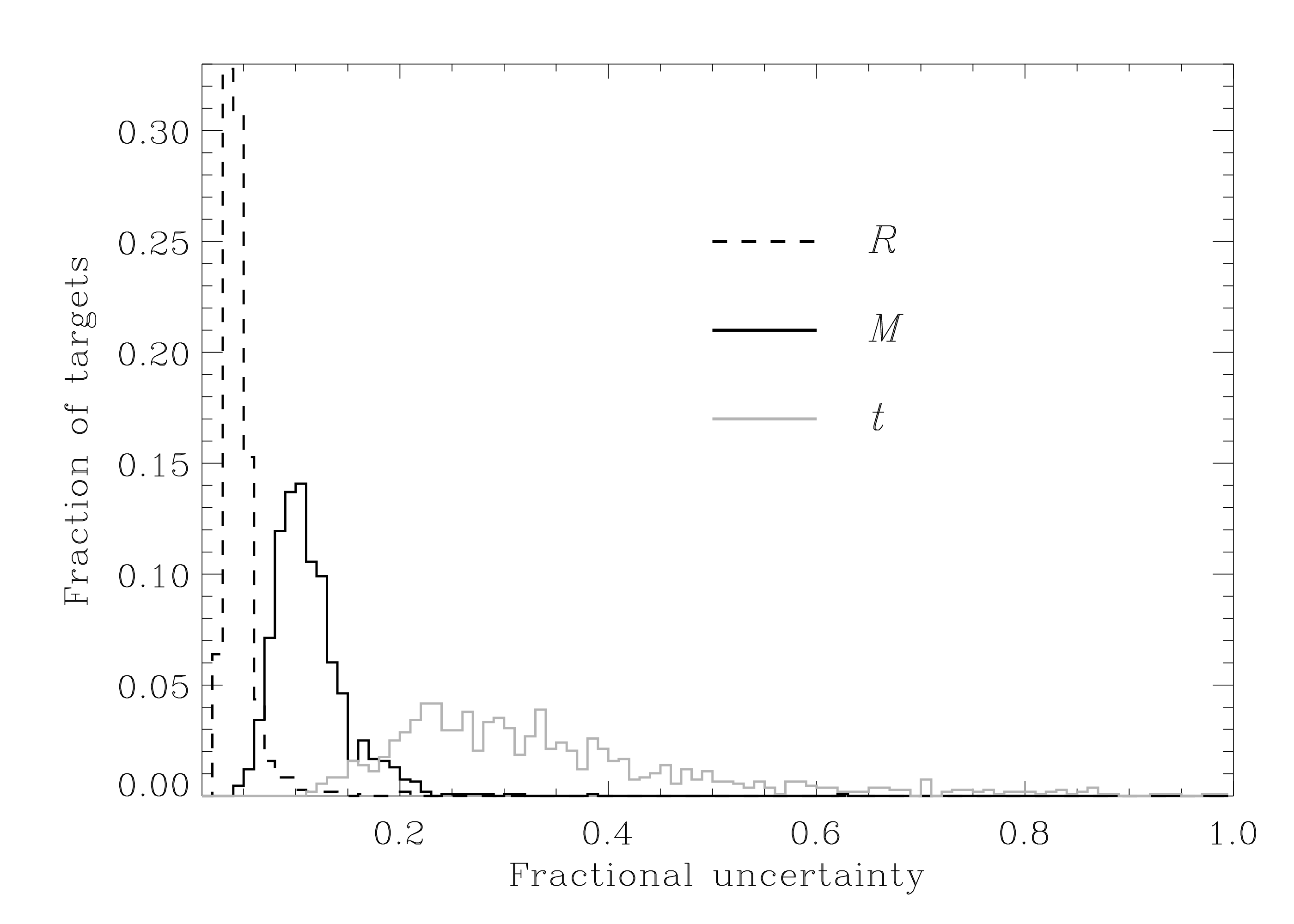}

\caption{Histograms of fractional uncertainties for estimated radii
  $R$, masses $M$, and ages $t$, of the full cohort of stars (using
  input effective temperatures from the IRFM set) (see figure
  legend).}

\label{fig:reshisto}
\end{figure*}

%%%%%%%%%%%%%%%%%%%%%%%%%%%%%%%%%%%%%%%%%%%%%%%%%%%%%%%%%%%%%%%%%%%%%%%

Whilst use of the seismic inputs \dnu\ and \numax\ has allowed much
tighter constraints to be placed on the ages of most of these stars
than would have been possible in the absence of such information, some
of the uncertainties are large. Figure~\ref{fig:reshisto} plots
histograms of the final (i.e., with the scatter contributions now
included) uncertainties on the ages, and also the masses and radii,
for the IRFM results. It will be possible to do much better for around
150 of these stars, by utilizing information from estimates of the
individual frequencies.

In sum: the ages presented here are not the bottom line on what is
possible using asteroseismology.

Nevertheless, ages have in most cases been estimated to a precision
that is useful in a statistical or ensemble sense. Just over 70\,\%
(SDSS results) and just under 60\,\% (IRFM results) of stars in the
full cohort have final fractional age uncertainties that are 30\,\% or
better. This fraction increases to over 80\,\% in the sample with
spectroscopic data, which only reinforces the point that to fully
utilize the diagnostic potential of the seismic data (or any other
data) for constraining the ages, good constraints on stellar
composition are required. It should of course be borne in mind that
the scatter found here will reflect model dependencies for the physics
adopted in the grids we used, and that other choices (which could
affect the ages) are possible.  The final, quoted uncertainties on the
ages presented here must be used/considered in any analysis: it is
these values that capture the systematics due to the stellar model
dependencies from the adopted physics in the commonly used grids we
have used, and differences in the adopted pipeline methodologies. It
should also be remembered that the ages come from one grid-pipeline
combination only, and age is the most model-dependent property.

Results from the Geneva Copenhagen Survey (GCS) provide a useful
comparison of what is possible for field stars in the absence of
results from seismology, when high-quality parallaxes, effective
temperatures and metallicities are available (Nordstr\"om et al.,
2004; Casagrande et al. 2011). An appropriate comparison is one made
with results from the Bruntt et al. cohort with spectrosopic data,
since metallicity information is available for all GCS targets. It
should be borne in mind that ages and masses from the GCS are based on
the use of two grids of models (Padova and BASTI), while in the
present paper systematics stemming from different models come from the
use of eleven grids.  The GCS sample is also brighter than the KASC
sample. Median age uncertainties for the Bruntt et al. cohort and the
CGS are quite similar.  As noted above, 80\,\% of Bruntt et al. cohort
have fractional ages uncertainties of 30\,\% or better.  However only
about 50\,\% of the stars in the GCS have errors $\le 30$\,\%.

\section{Conclusion}
\label{sec:conc}

We have presented a homogeneous asteroseismic analysis of over 500
solar-type main-sequence and sub-giant stars observed by
\emph{Kepler}. Stellar properties were estimated using two global
asteroseismic parameters---the average large frequency separation,
\dnu, and the frequency of maximum oscillations power, \numax---and
complementary photometric and spectroscopic data. Homogeneous sets of
effective temperatures, $T_{\rm eff}$, were available for the entire
ensemble using available complementary photometry; spectroscopic
estimates of $T_{\rm eff}$ and [Fe/H] were available from a
homogeneous analysis of ground-based data on a subset totalling 87
stars.

We provide estimates of stellar radii, masses, surface gravities, mean
densities and ages. We add a strong note of caution regarding the
ages. The quoted age uncertainties are similar to those obtained in
the best case scenario from isochrone fitting of field stars when
parallaxes, metallicities and effective temperatures are known
precisely (e.g., Nordstrom et al. 2004; Casagrande et
al. 2011). Although the advantages and the potential of
asteroseismology are obvious, many of the usual warnings thus apply to
the present sample. The ages in this paper will be useful for ensemble
studies, but should be treated carefully when considered on a
star-by-star basis. Because we used only two global asteroseismic
parameters, future analyses using individual oscillation
frequencies---which are important for asteroseismic estimation of
ages---will offer significant improvements on up to 150 stars.

A future priority will be to collect further high-quality spectroscopy
and photometry in appropriate colours (e.g., the Str\"omgren filters),
in order to place tight constraints on the compositions of all stars
in the sample. This should be possible from both episodic ground-based
campaigns, and from utilizing data from planned spectroscopic surveys,
e.g., APOGEE (Majewski et al.  2010). GAIA (see e.g., Gilmore et
al. 2012) will provide exquisite parallaxes on the cohort of stars in
this paper. Since the asteroseismic data can provide accurate surface
gravities, they can also play an important role in helping to
calibrate spectroscopic analyses. A formal collaboration (APOKASC) has
already been established between APOGEE and the \emph{Kepler}
Asteroseismic Science Consortium, with the full sample discussed in
this paper having already been included in target planning.

\acknowledgements 

We thank the anonymous referee for helpful comments that improved the
paper. Funding for this Discovery mission is provided by NASA's
Science Mission Directorate. The authors wish to thank the entire
\emph{Kepler} team, without whom these results would not be possible.
W.J.C. acknowledges the support of the UK Science and Technology
Facilities Council (STFC). S.B. ackowledges support from NSF grant
AST-1105930 and NASA grant NNX13AE70G.  D.H. is supported by an
appointment to the NASA Postdoctoral Program at Ames Research Center,
administered by Oak Ridge Associated Universities through a contract
with NASA. A.M.S is partially supported by the International
Reintegration Grant PIRG-GA-2009-247732, and the MICINN grant
AYA2011-24704.  L.G., W.H.B., and J.P.M. acknowledge support from
Collaborative Research Center CRC 963 ``Astrophysical Flow
Instabilities and Turbulence" (Project A18), funded by the German
Research Foundation (DFG).  O.L.C. was a Henri Poincar\'e Fellow at
the Observatoire de la C\^ote d'Azur. The Henri Poincar\'e Fellowship
is funded by the Conseil G\'en\'eral des Alpes-Maritimes and the
Observatoire de la C\^ote d'Azur. J.M-\.Z acknowledges support from
MNiSW grant N\,N203\,405139. T.S.M. acknowledges NASA grant
NNX13AE91G. R.A.G. acknowledges the funding from the European
Community's Seventh Framework Programme (FP7/2007-2013) under grant
agreement No. 269194 (IRSES/ASK). R.A.G. and B.M. acknowledge support
from the ANR program IDEE ``Interaction Des Etoiles et des
Exoplan\`etes" (Agence Nationale de la Recherche, France).
S.H. acknowledges financial support from the Netherlands Organisation
for Scientific Research (NWO).

Funding for the Stellar Astrophysics Centre is provided by The Danish
National Research Foundation (Grant agreement no.: DNRF106). The
research is supported by the ASTERISK project (ASTERoseismic
Investigations with SONG and Kepler) funded by the European Research
Council (Grant agreement no.: 267864). The research leading to these
results has received funding from the European Community's Seventh
Framework Programme ([FP7/2007-2013]) under grant agreement no. 312844
(SPACEINN).

We are also grateful for support from the International Space Science
Institute (ISSI).

\clearpage

%%%%%%%%%%%%%%%%%%%%%%%%%%%%%%%%%%%%%%%%%%%%%%%%%%%%%%%%%%%%%%%%%%%%%%%%%%%%%%%%%%%%%%%

%   $   0 \pm    0$  $     ...     $

\begin{deluxetable}{cccccc}

\tabletypesize{\scriptsize} \tablecaption{Seismic and non-seismic
  input parameters for the full cohort, with photometric
  SDSS-calibrated and IRFM $T_{\rm eff}$ and field-average [Fe/H].}
\tablewidth{0pt}
\tablehead{
 \colhead{KIC}&
 \colhead{\numax}& \colhead{\dnu}&
 \colhead{SDSS $T_{\rm eff}$}&
 \colhead{IRFM $T_{\rm eff}$}& \colhead{[Fe/H]}\\ 
 \colhead{}&
 \colhead{($\rm \mu Hz$)}& \colhead{($\rm \mu Hz$)}& \colhead{(K)}& \colhead{(K)}& \colhead{(dex)}}
 \startdata
  1430163& $1867 \pm   92$& $ 84.6 \pm   2.0$& $6796 \pm   78$& $6806 \pm  177$& $-0.20 \pm 0.30$\\
  1435467& $1295 \pm   52$& $ 70.8 \pm   0.8$& $6433 \pm   86$& $6521 \pm  164$& $-0.20 \pm 0.30$\\
  1725815& $1045 \pm   47$& $ 55.4 \pm   1.3$& $6550 \pm   82$& $6532 \pm  165$& $-0.20 \pm 0.30$\\
  2010607& $ 675 \pm   86$& $ 42.5 \pm   1.7$& $6361 \pm   71$& $6796 \pm  175$& $-0.20 \pm 0.30$\\
  2309595& $ 643 \pm   20$& $ 39.3 \pm   2.2$& $5238 \pm   65$& $5315 \pm  112$& $-0.20 \pm 0.30$\\
  2450729& $1053 \pm   68$& $ 61.9 \pm   0.6$& $6029 \pm   59$& $6093 \pm  152$& $-0.20 \pm 0.30$\\
  2837475& $1630 \pm   54$& $ 75.1 \pm   1.3$& $6688 \pm   57$& $6715 \pm  174$& $-0.20 \pm 0.30$\\
  2849125& $ 728 \pm   27$& $ 41.4 \pm   1.8$& $6158 \pm   56$& $6421 \pm  154$& $-0.20 \pm 0.30$\\
\enddata
 \label{tab:data1}
\end{deluxetable}

%%%%%%%%%%%%%%%%%%%%%%%%%%%%%%%%%%%%%%%%%%%%%%%%%%%%%%%%%%%%%%%%%%%%%%%%%%%%%%%%%%%%%%%

\begin{deluxetable}{ccccc}
\tabletypesize{\scriptsize} \tablecaption{Seismic and non-seismic
  input parameters for the sample with spectroscopic $T_{\rm eff}$ and
  [Fe/H] from Bruntt et al. (2012)}
\tablewidth{0pt}
\tablehead{
 \colhead{KIC}&
 \colhead{\numax}& \colhead{\dnu}&
 \colhead{$T_{\rm eff}$}&
 \colhead{[Fe/H]}\\ 
 \colhead{}&
 \colhead{($\rm \mu Hz$)}& \colhead{($\rm \mu Hz$)}& \colhead{(K)}& \colhead{(dex)}}
 \startdata
  1430163& $1867 \pm   92$& $ 84.6 \pm   2.0$& $6520 \pm   84$& $-0.11 \pm 0.09$\\
  1435467& $1295 \pm   52$& $ 70.8 \pm   0.8$& $6264 \pm   84$& $-0.01 \pm 0.09$\\
  2837475& $1630 \pm   54$& $ 75.1 \pm   1.3$& $6700 \pm   84$& $-0.02 \pm 0.09$\\
  3424541& $ 745 \pm   55$& $ 41.5 \pm   1.1$& $6080 \pm   84$& $ 0.01 \pm 0.09$\\
  3427720& $2756 \pm  191$& $119.9 \pm   2.0$& $6040 \pm   84$& $-0.03 \pm 0.09$\\
  3456181& $ 921 \pm   30$& $ 52.0 \pm   0.8$& $6270 \pm   84$& $-0.19 \pm 0.09$\\
  3632418& $1159 \pm   44$& $ 60.9 \pm   0.4$& $6190 \pm   84$& $-0.16 \pm 0.09$\\
  3656476& $1887 \pm   40$& $ 93.3 \pm   1.3$& $5710 \pm   84$& $ 0.34 \pm 0.09$\\
\enddata
 \label{tab:data2}
\end{deluxetable}

%%%%%%%%%%%%%%%%%%%%%%%%%%%%%%%%%%%%%%%%%%%%%%%%%%%%%%%%%%%%%%%%%%%%%%%%%%%%%%%%%%%%%%%

%%%%%%%%%%%%%%%%%%%%%%%%%%%%%%%%%%%%%%%%%%%%%%%%%%%%%%%%%%%%%%%%%%%%%%%%%%%%%%%%%%%%%%%

\begin{deluxetable}{lcclclllcccc}
\tabletypesize{\tiny} 
\rotate
\tablecaption{Grid-based pipelines and the grids of models} 
\tablewidth{0pt}
\tablehead{
 \colhead{Pipeline}&
 \colhead{Grid}&
 \colhead{Mixing}&
 \colhead{Definition}&
 \colhead{Diffusion}&
 \colhead{Core}&
 \colhead{$Y_0$}&
 \colhead{Enrichment}&
 \colhead{Reaction}&
 \colhead{High/low-$T$}&
 \colhead{$Z$-Mixture}&
 \colhead{EOS}\\
 \colhead{}&
 \colhead{}&
 \colhead{Length $\alpha$}&
 \colhead{[Fe/H]$=0$}&
 \colhead{}&
 \colhead{overshoot}&
 \colhead{}&
 \colhead{law}&
 \colhead{rates}&
 \colhead{opacities}&
 \colhead{}&
 \colhead{} }
 \startdata

 YB& Dartmouth& 1.938& $Z/X=0.023$& $Y$, $Z$& Func. of $M, Z/X$& variable& $Y_0=0.245+1.54Z_0$& A98$^{(1)}$ & OPAL/F05& GS98& Ideal$^{(2)}$\\

 YB& Marigo& 1.680& $Z=0.019$& None& Func. of $M$& variable& $Y=0.23+2.25Z$  & CF88/LP90 & OPAL/AF& GN93& MHD/ST88  \\

 YB& YY& 1.7432& $Z/X=0.0244$& $Y$& Func. of $M$& variable&  $Y_0=0.23+2Z_0$& B89& OPAL/AF& GN93& OPAL96 \\

 YB& YREC& 1.826& $Z/X=0.023$& $Y$, $Z$& $0.2H_p$& variable$^{(3)}$&  $\Delta Y / \Delta Z =1$& NACRE& OPAL/F05& GS98& OPAL05 \\

 YB& YREC2& 1.4 to 2.2& $Z/X=0.023$& None & $0.2H_p$& variable& $Y_0=0.245+1.54Z_0$& A98$^{(4)}$& OPAL/F05& AGSS09& OPAL05 \\

 BeSPP & GARSTEC& 1.81& $Z/X=0.0229$& $Y$, $Z$& Diffusive$^{(5)}$& variable$^{(6)}$& $Y_0=Y_i+1.4(Z-Z_i)$& A11& OPAL/F05& GS98& C03 \\

 BeSPP & BASTI& 1.913  &$Z/X=0.0245$& None  & None  & variable$^{(7)}$    & $\Delta Y/\Delta Z=1.4$& NACRE$^{(8)}$   &  OPAL/AF   & GN93    & C03       \\

 RadEx10& ASTEC1& 2.0& $Z/X=0.02456$& None& $0.25H_p$& variable$^{(9)}$& $Y=0.29-Z$ & BP95& OPAL/K91& GN93& E73 \\
\enddata

\tablecomments{\\
               (1): Except N($p,\gamma$)O, $3\alpha$, C($\alpha,\gamma$)O;
               (2): ideal EOS with Debye-Huckle correction ($M \ge 0.8\,\rm M_{\odot}$);
               (3): $Y_0=0.246$ at $Z_0=0.01695$;
               (4): except for N($p,\gamma$)O\\
               (5): diffusive ($f=0.016$) but restricted to smaller values for small convective cores;
               (6): $Z_i=0.01876$, $Y_i=0.26896$;
               (7): $Y_0=245$ at $Z_0=0.0001$;
               (8): except for C($\alpha,\gamma$)O.\\
               (9): $X_i$ kept constant at 0.71 for different $Z_i$;
               (10): $X_i$ kept constant at 0.7,$Z_i$ varies; 
               (11): Different combinations of $Z_i$ and $X_i$. Many $X_i$ for same $Z_i$;\\
               (12): $Z/Z_{\odot}$ from 0.5 to 2.0, where $Z_{\odot}=0.13757$;  masses from $0.75M_{\odot}$ to $1.75M_{\odot}$
               {\ }\\
               A98: Adelberger et al. (1998);
               A11: Adelberger et al. (2011);
               AF94: Alexander \& Ferguson (1994);
               AGS05: Asplund et al. (2005);
               AGSS09: Asplund et al. (2009);\\
               B89: Bahcall (1989);
               BP95: Bahcall \& Pinsonneault (1995);
               CF88: Caughlan \& Fowler (1988);
               C03: Cassisi et al. (2003);
               E73: Eggleton et al. (1973);\\
               F05: Ferguson et al. (2005);
               G11: Gai et al. (2011);
               GN93: Grevesse \& Noels (1993);
               GS98: Grevesse \& Sauval (1998);\\
               K91: Kurucz et al. (1991);
               LP90: Landr\'e et al. (1990);
               MHD: Mihalas et al. (1988);
               ST88: Straneiro (1988)\\
}
                
\label{tab:pipelines}
\end{deluxetable}

%%%%%%%%%%%%%%%%%%%%%%%%%%%%%%%%%%%%%%%%%%%%%%%%%%%%%%%%%%%%%%%%%%%%%%%%%%%%%%%%%%%%%%%

%%%%%%%%%%%%%%%%%%%%%%%%%%%%%%%%%%%%%%%%%%%%%%%%%%%%%%%%%%%%%%%%%%%%%%%%%%%%%%%%%%%%%%%

\begin{deluxetable}{cccccccccccccccc}
\tabletypesize{\scriptsize} \tablecaption{Estimated stellar properties
  using SDSS-calibrated $T_{\rm eff}$ and field-average [Fe/H]
  values.}
\tablewidth{0pt}
\tablehead{
 \colhead{KIC}&
 \colhead{$M$}& 
 \colhead{$R$}&
 \colhead{$\rho$}&
 \colhead{$\log\,g$}&
 \colhead{$t$}\\
 \colhead{}&
 \colhead{($\rm M_{\odot}$)}&
 \colhead{($\rm R_{\odot}$)}&
 \colhead{($\rm \rho_{\odot}$)}&
 \colhead{(dex)}&
 \colhead{(Gyr)}}
 \startdata
  1430163& $  1.38\,+  0.14\,-  0.10$& $  1.49\,+  0.05\,-  0.05$& $0.4195\,+0.0189\,-0.0181$& $ 4.234\,+ 0.014\,- 0.015$& $ 1.3\,+ 0.6\,- 0.7$\\
  1435467& $  1.16\,+  0.20\,-  0.06$& $  1.63\,+  0.08\,-  0.05$& $0.2732\,+0.0073\,-0.0071$& $ 4.088\,+ 0.019\,- 0.014$& $ 4.7\,+ 0.7\,- 1.7$\\
  1725815& $  1.43\,+  0.15\,-  0.11$& $  2.02\,+  0.08\,-  0.08$& $0.1732\,+0.0077\,-0.0074$& $ 3.982\,+ 0.014\,- 0.014$& $ 3.1\,+ 0.7\,- 0.8$\\
  2010607& $  1.37\,+  0.15\,-  0.14$& $  2.42\,+  0.12\,-  0.11$& $0.0972\,+0.0079\,-0.0075$& $ 3.809\,+ 0.025\,- 0.025$& $ 3.8\,+ 0.8\,- 0.8$\\
  2309595& $  1.14\,+  0.19\,-  0.22$& $  2.40\,+  0.19\,-  0.23$& $0.0828\,+0.0078\,-0.0067$& $ 3.734\,+ 0.013\,- 0.013$& $ 6.0\,+ 4.5\,- 1.5$\\
  2450729& $  1.11\,+  0.12\,-  0.13$& $  1.76\,+  0.07\,-  0.07$& $0.2017\,+0.0054\,-0.0052$& $ 3.988\,+ 0.017\,- 0.017$& $ 6.2\,+ 1.8\,- 1.2$\\
  2837475& $  1.42\,+  0.14\,-  0.10$& $  1.63\,+  0.06\,-  0.05$& $0.3286\,+0.0107\,-0.0107$& $ 4.168\,+ 0.012\,- 0.012$& $ 1.9\,+ 0.4\,- 0.7$\\
  2849125& $  1.38\,+  0.29\,-  0.12$& $  2.42\,+  0.18\,-  0.10$& $0.0975\,+0.0061\,-0.0058$& $ 3.814\,+ 0.016\,- 0.016$& $ 4.0\,+ 0.7\,- 1.6$\\
\enddata
 \label{tab:res1}
\end{deluxetable}

%%%%%%%%%%%%%%%%%%%%%%%%%%%%%%%%%%%%%%%%%%%%%%%%%%%%%%%%%%%%%%%%%%%%%%%%%%%%%%%%%%%%%%%

\begin{deluxetable}{cccccccccccccccc}
\tabletypesize{\scriptsize} \tablecaption{Estimated stellar properties
  using IRFM $T_{\rm eff}$ and field-average [Fe/H] values.}
\tablewidth{0pt}
\tablehead{
 \colhead{KIC}&
 \colhead{$M$}& 
 \colhead{$R$}&
 \colhead{$\rho$}&
 \colhead{$\log\,g$}&
 \colhead{$t$}\\
 \colhead{}&
 \colhead{($\rm M_{\odot}$)}&
 \colhead{($\rm R_{\odot}$)}&
 \colhead{($\rm \rho_{\odot}$)}&
 \colhead{(dex)}&
 \colhead{(Gyr)}}
 \startdata
  1430163& $  1.38\,+  0.15\,-  0.09$& $  1.49\,+  0.05\,-  0.05$& $0.4219\,+0.0199\,-0.0198$& $ 4.237\,+ 0.017\,- 0.017$& $ 1.3\,+ 0.6\,- 0.9$\\
  1435467& $  1.19\,+  0.20\,-  0.08$& $  1.64\,+  0.08\,-  0.05$& $0.2750\,+0.0082\,-0.0081$& $ 4.091\,+ 0.022\,- 0.016$& $ 4.5\,+ 0.8\,- 1.7$\\
  1725815& $  1.43\,+  0.15\,-  0.13$& $  2.02\,+  0.08\,-  0.08$& $0.1733\,+0.0081\,-0.0077$& $ 3.982\,+ 0.016\,- 0.016$& $ 3.2\,+ 0.9\,- 0.9$\\
  2010607& $  1.48\,+  0.15\,-  0.14$& $  2.44\,+  0.12\,-  0.11$& $0.1022\,+0.0086\,-0.0083$& $ 3.835\,+ 0.027\,- 0.028$& $ 3.1\,+ 0.8\,- 0.8$\\
  2309595& $  1.18\,+  0.19\,-  0.22$& $  2.42\,+  0.18\,-  0.21$& $0.0825\,+0.0072\,-0.0063$& $ 3.737\,+ 0.013\,- 0.014$& $ 5.5\,+ 3.8\,- 1.2$\\
  2450729& $  1.13\,+  0.11\,-  0.13$& $  1.77\,+  0.06\,-  0.07$& $0.2026\,+0.0055\,-0.0054$& $ 3.992\,+ 0.018\,- 0.017$& $ 5.8\,+ 1.9\,- 1.2$\\
  2837475& $  1.47\,+  0.15\,-  0.13$& $  1.64\,+  0.06\,-  0.06$& $0.3325\,+0.0134\,-0.0127$& $ 4.173\,+ 0.020\,- 0.016$& $ 1.8\,+ 0.6\,- 0.9$\\
  2849125& $  1.52\,+  0.17\,-  0.19$& $  2.47\,+  0.15\,-  0.14$& $0.0988\,+0.0062\,-0.0060$& $ 3.826\,+ 0.016\,- 0.016$& $ 2.9\,+ 1.2\,- 0.8$\\
\enddata
 \label{tab:res2}
\end{deluxetable}

%%%%%%%%%%%%%%%%%%%%%%%%%%%%%%%%%%%%%%%%%%%%%%%%%%%%%%%%%%%%%%%%%%%%%%%%%%%%%%%%%%%%%%%

\begin{deluxetable}{cccccccccccccccc}
\tabletypesize{\scriptsize} \tablecaption{Estimated stellar properties
  using spectroscopic $T_{\rm eff}$ and [Fe/H] values from Bruntt et
  al. (2012).}
\tablewidth{0pt}
\tablehead{
 \colhead{KIC}&
 \colhead{$M$}& 
 \colhead{$R$}&
 \colhead{$\rho$}&
 \colhead{$\log\,g$}&
 \colhead{$t$}\\
 \colhead{}&
 \colhead{($\rm M_{\odot}$)}&
 \colhead{($\rm R_{\odot}$)}&
 \colhead{($\rm \rho_{\odot}$)}&
 \colhead{(dex)}&
 \colhead{(Gyr)}}
 \startdata
  1430163& $  1.34\,+  0.06\,-  0.06$& $  1.48\,+  0.03\,-  0.03$& $0.4099\,+0.0171\,-0.0174$& $ 4.221\,+ 0.013\,- 0.014$& $ 1.9\,+ 0.6\,- 0.5$\\
  1435467& $  1.31\,+  0.08\,-  0.16$& $  1.69\,+  0.04\,-  0.06$& $0.2675\,+0.0066\,-0.0067$& $ 4.094\,+ 0.012\,- 0.019$& $ 3.7\,+ 1.7\,- 0.8$\\
  2837475& $  1.41\,+  0.06\,-  0.04$& $  1.63\,+  0.03\,-  0.03$& $0.3295\,+0.0105\,-0.0103$& $ 4.167\,+ 0.010\,- 0.010$& $ 2.0\,+ 0.3\,- 0.4$\\
  3424541& $  1.39\,+  0.07\,-  0.06$& $  2.47\,+  0.06\,-  0.06$& $0.0930\,+0.0046\,-0.0044$& $ 3.797\,+ 0.016\,- 0.015$& $ 4.0\,+ 0.6\,- 0.6$\\
  3427720& $  1.08\,+  0.06\,-  0.06$& $  1.12\,+  0.02\,-  0.02$& $0.7817\,+0.0262\,-0.0255$& $ 4.378\,+ 0.012\,- 0.012$& $ 3.5\,+ 1.5\,- 1.4$\\
  3456181& $  1.36\,+  0.07\,-  0.12$& $  2.10\,+  0.05\,-  0.06$& $0.1463\,+0.0042\,-0.0040$& $ 3.924\,+ 0.010\,- 0.012$& $ 3.9\,+ 1.0\,- 0.5$\\
  3632418& $  1.31\,+  0.08\,-  0.07$& $  1.88\,+  0.04\,-  0.04$& $0.1993\,+0.0034\,-0.0034$& $ 4.010\,+ 0.010\,- 0.009$& $ 4.1\,+ 0.5\,- 0.6$\\
  3656476& $  1.10\,+  0.05\,-  0.06$& $  1.34\,+  0.03\,-  0.03$& $0.4588\,+0.0109\,-0.0105$& $ 4.225\,+ 0.008\,- 0.008$& $ 7.2\,+ 2.2\,- 1.8$\\
\enddata
 \label{tab:res3}
\end{deluxetable}

%%%%%%%%%%%%%%%%%%%%%%%%%%%%%%%%%%%%%%%%%%%%%%%%%%%%%%%%%%%%%%%%%%%%%%%%%%%%%%%%%%%%%%%

\clearpage

\appendix

\section{Comparison plots for different grid-pipeline combinations}
\label{sec:app}

\begin{figure*}
\epsscale{1.0}
\plotone{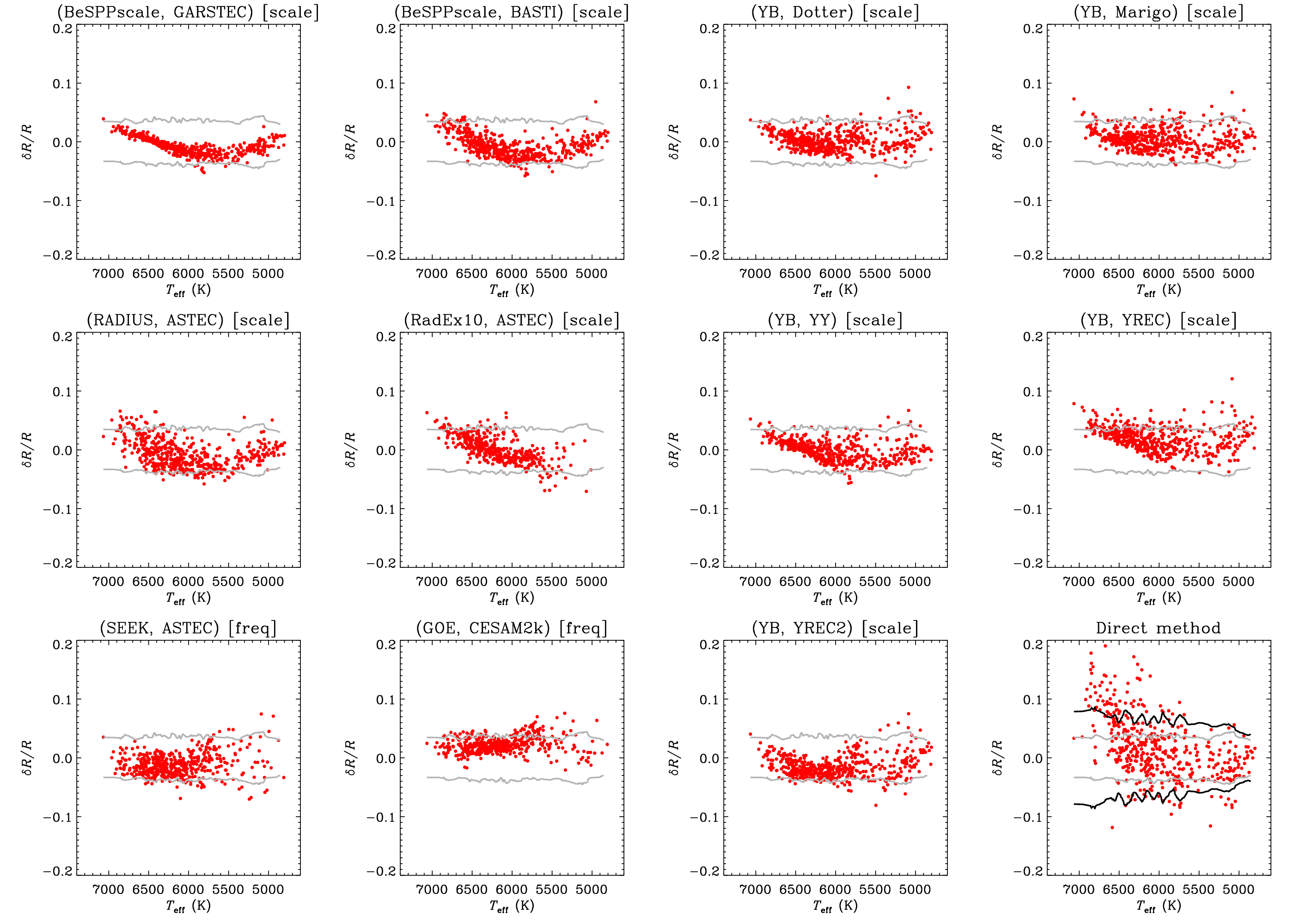}

\caption{Fractional differences in estimated radii, $R$, for analyses
  performed on the entire ensemble, with \dnu\ and $\nu_{\rm max}$,
  the photometric (IRFM) $T_{\rm eff}$ and field [Fe/H] values used as
  inputs. The plots show differences with respect to the BeSPP
  pipeline run with the GARSTEC grid run using model-calculated
  eigenfrequencies to estimate the \dnu\ of each model in its
  grid. Gray lines mark the median $1\sigma$ envelope of the
  grid-pipeline returned, formal uncertainties.  These lines are
  included to help judge the \emph{typical} precision only. The bottom
  right-hand panel shows results from direct application of the
  scaling relations, the black lines showing the median $1\sigma$
  envelope on the resulting uncertainties.}
\label{fig:diffteff1_a}
\end{figure*}

\begin{figure*}
\epsscale{1.0}
\plotone{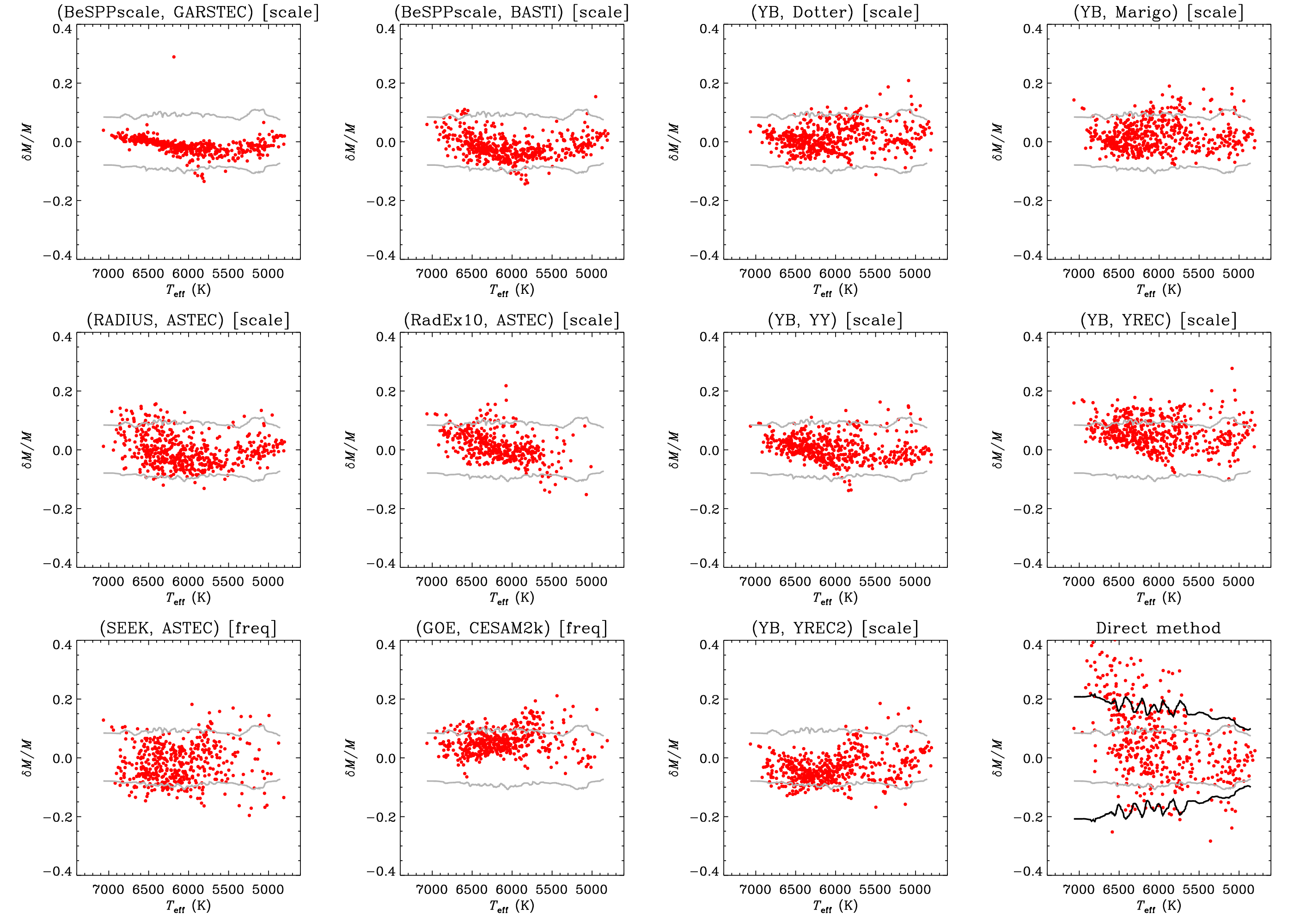}
\caption{As per Figure~\ref{fig:diffteff1_a}, but for fractional
  differences in mass $M$.}
\label{fig:diffteff1_b}
\end{figure*}

\begin{figure*}
\epsscale{1.0}
\plotone{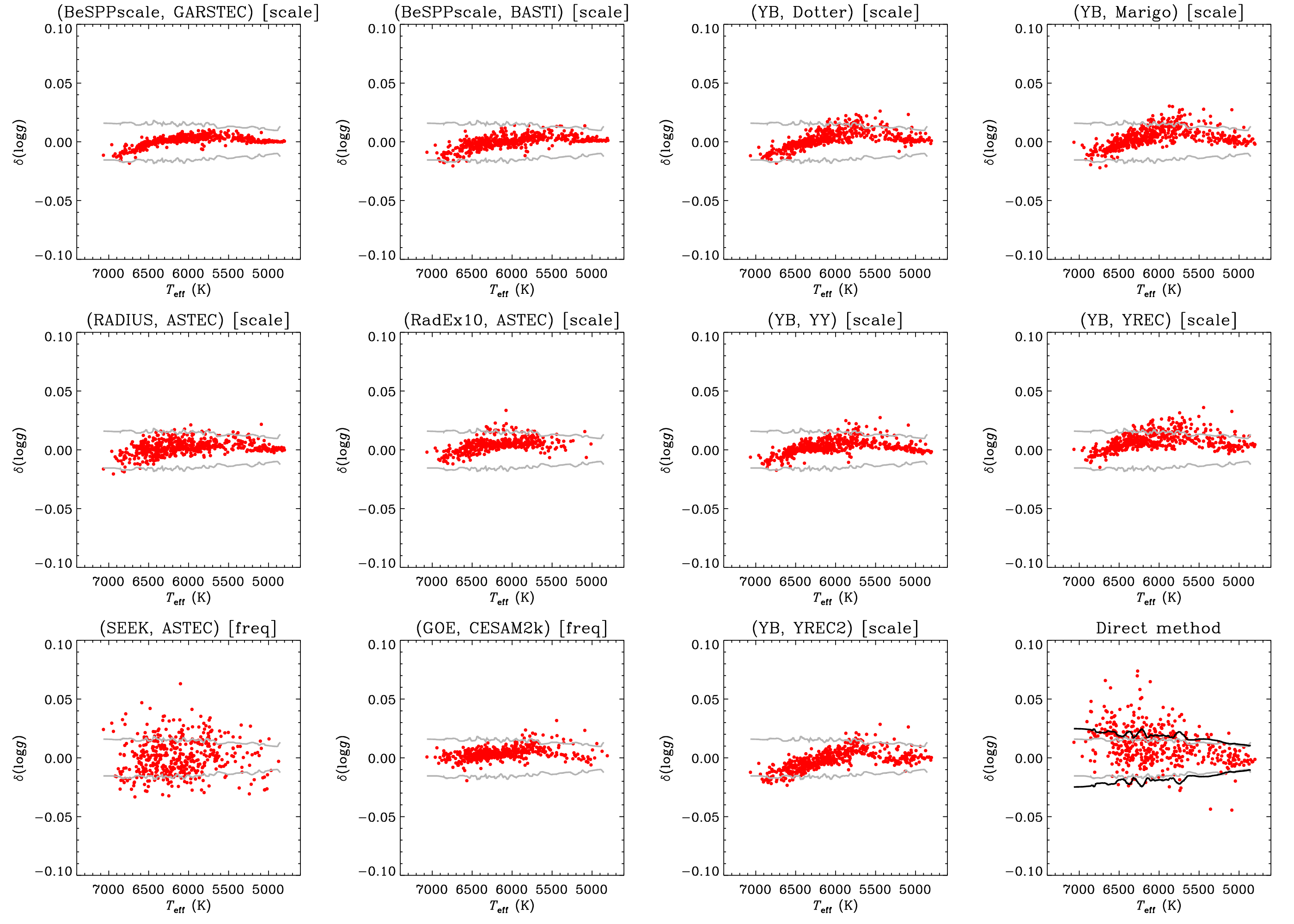}
\caption{As per Figure~\ref{fig:diffteff1_a}, but for differences in
  $\log\,g$.}
\label{fig:diffteff1_c}
\end{figure*}

\begin{figure*}
\epsscale{1.0}
\plotone{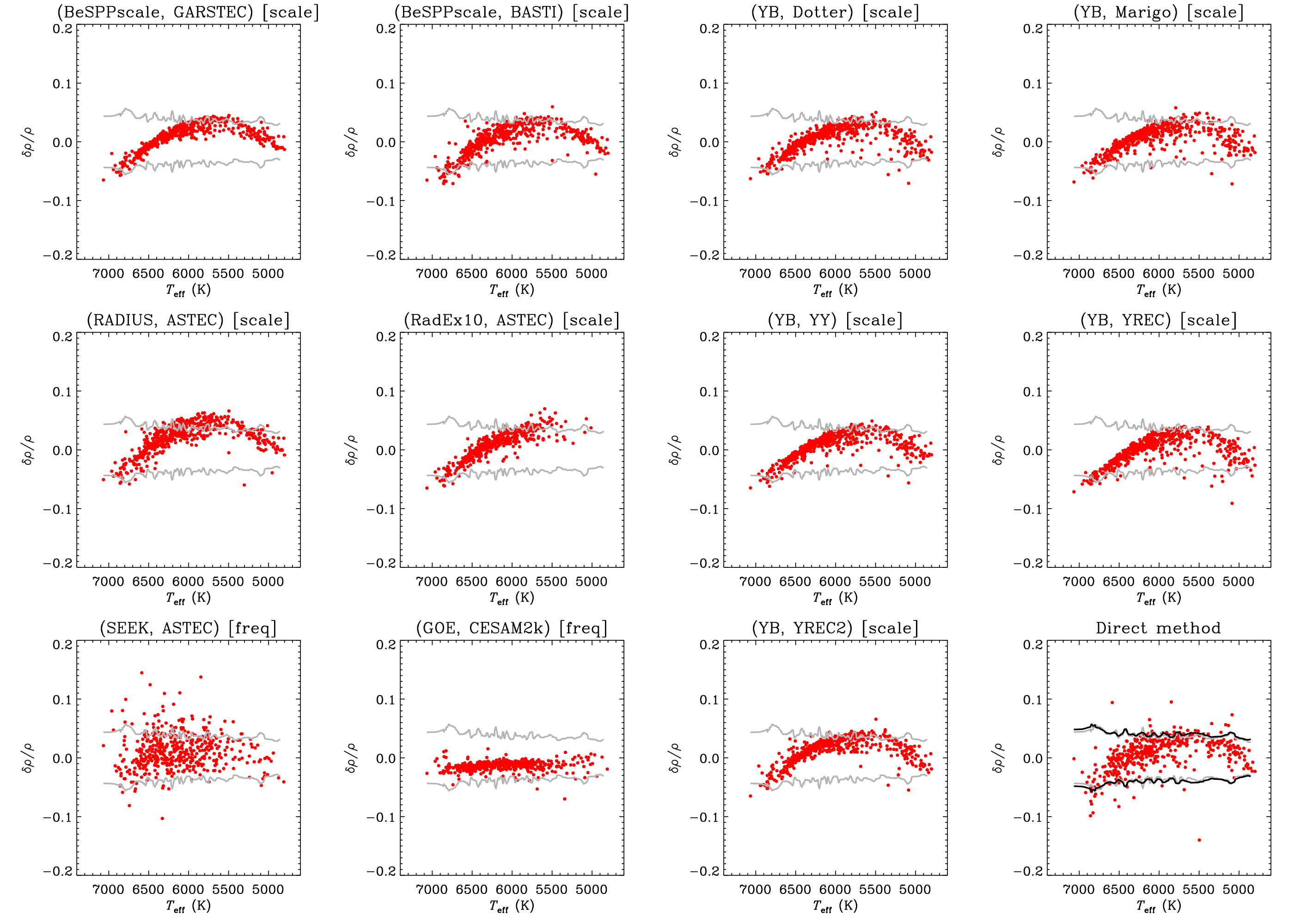}
\caption{As per Figure~\ref{fig:diffteff1_a}, but for fractional
  differences in average density $\rho$.}
\label{fig:diffteff1_d}
\end{figure*}

\begin{figure*}
\epsscale{1.0}
\plotone{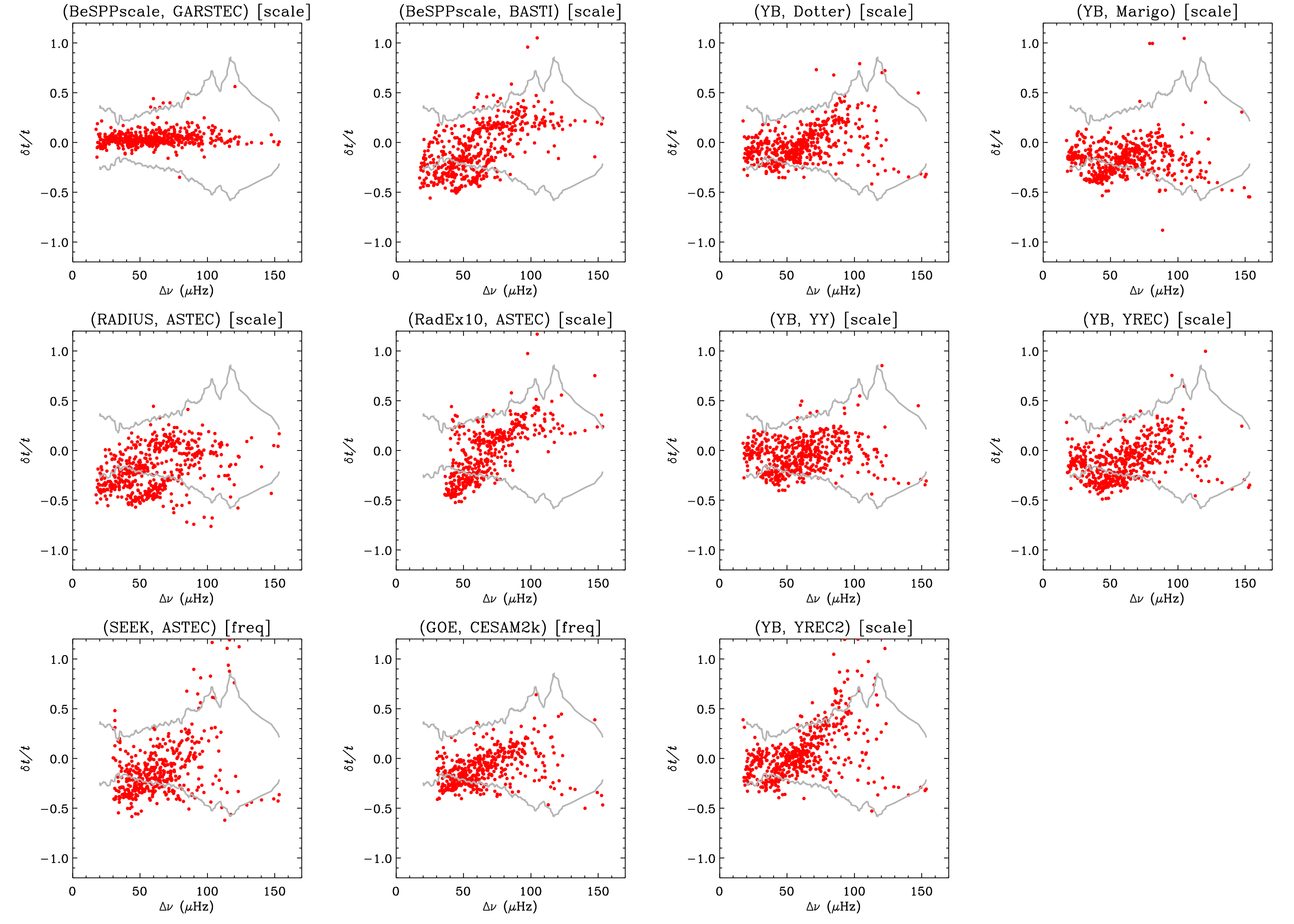}
\caption{As per Figure~\ref{fig:diffteff1_a}, but for fractional
  differences in age, $t$, and with \dnu\ as the
  independent variable for the plot.}
\label{fig:diffteff1_e}
\end{figure*}

Figures~\ref{fig:diffteff1_a} through~\ref{fig:diffteff1_e} show
comparisons of results from the different grid-pipeline combinations,
for analyses performed on the entire ensemble, with \dnu, $\nu_{\rm
  max}$, the photometric (IRFM) $T_{\rm eff}$, and field [Fe/H] values
used as inputs. The plots show differences with respect to a common
reference set of results, those given by BeSPP using the GARSTEC grid
and model-calculated eigenfrequencies to estimate the \dnu\ of each
model (the ``frequency'' mode of this pipeline).

The plotted differences are fractional differences in $R$, $M$, $\rho$
and age $t$; and absolute differences in $\log\,g$. The gray lines
mark envelopes corresponding to the median of the $1\sigma$
uncertainties returned by all grid pipelines, with medians calculated
in 10-target batches sorted on the independent variable used for the
plots ($T_{\rm eff}$ for $R$, $M$, $\log\,g$ and $\rho$; and \dnu\ for
$t$). The results shown in the lower right-hand plots are those given
by direct application of the scaling relations (note there is no
direct-method estimate of age $t$). The thick black lines mark the
$1\sigma$ uncertainty envelopes for the direct method. We note that
SEEK, RadEx10 and GOE were coupled to grids whose sampling was less
dense for more evolved (generally lower $T_{\rm eff}$) stars: hence,
for those cases some results have not been returned on targets in this
part of the parameter space.

The direct-method results are typically more scattered, but on the
whole consistent with the larger uncertainties expected from basic
error-propagation. The uncertainties are smaller for the grid-based
searches because the solutions are constrained to satisfy stellar
evolution theory, hence only a narrow range of outcomes is permitted.
The native fractional uncertainties on $\nu_{\rm max}$ are, on
average, about twice the size of those on \dnu.  This acts to drive up
the direct-method uncertainties, most notably on $R$ and $M$. The
direct-method uncertainties for $\rho$, which are determined solely by
uncertainties in \dnu, are in contrast seen to match quite closely the
uncertainties given by the grid-based searches.

\begin{figure*}
\epsscale{1.0}
\plotone{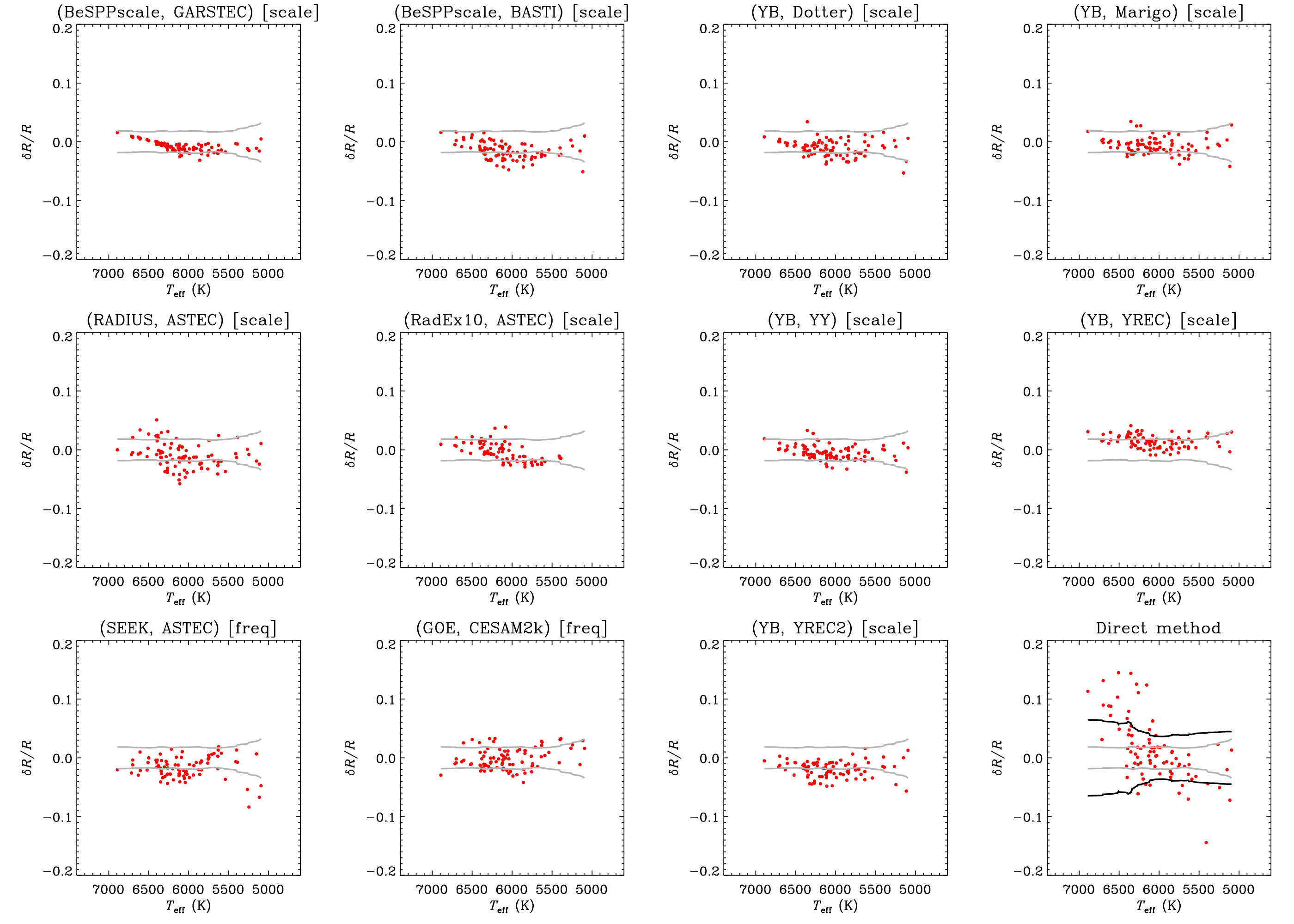}

\caption{Fractional differences in estimated radii, $R$, for analyses
  performed on a subset 89 stars using spectroscopic $T_{\rm eff}$ and
  [Fe/H] from Bruntt et al. (2012) as inputs, along with with
  \dnu\ and $\nu_{\rm max}$. Plots again show differences with respect
  to the BeSPP pipeline run with the GARSTEC grid, run using
  model-calculated eigenfrequencies to estimate the \dnu\ of each
  model in its grid. Gray lines mark the median $1\sigma$ envelope of
  the returned, formal uncertainties. The bottom right-hand panel
  shows results from direct application of the scaling relations, the
  black lines showing the median $1\sigma$ envelope on the resulting
  uncertainties.}

\label{fig:diffteff2_a}
\end{figure*}

\begin{figure*}
\epsscale{1.0}
\plotone{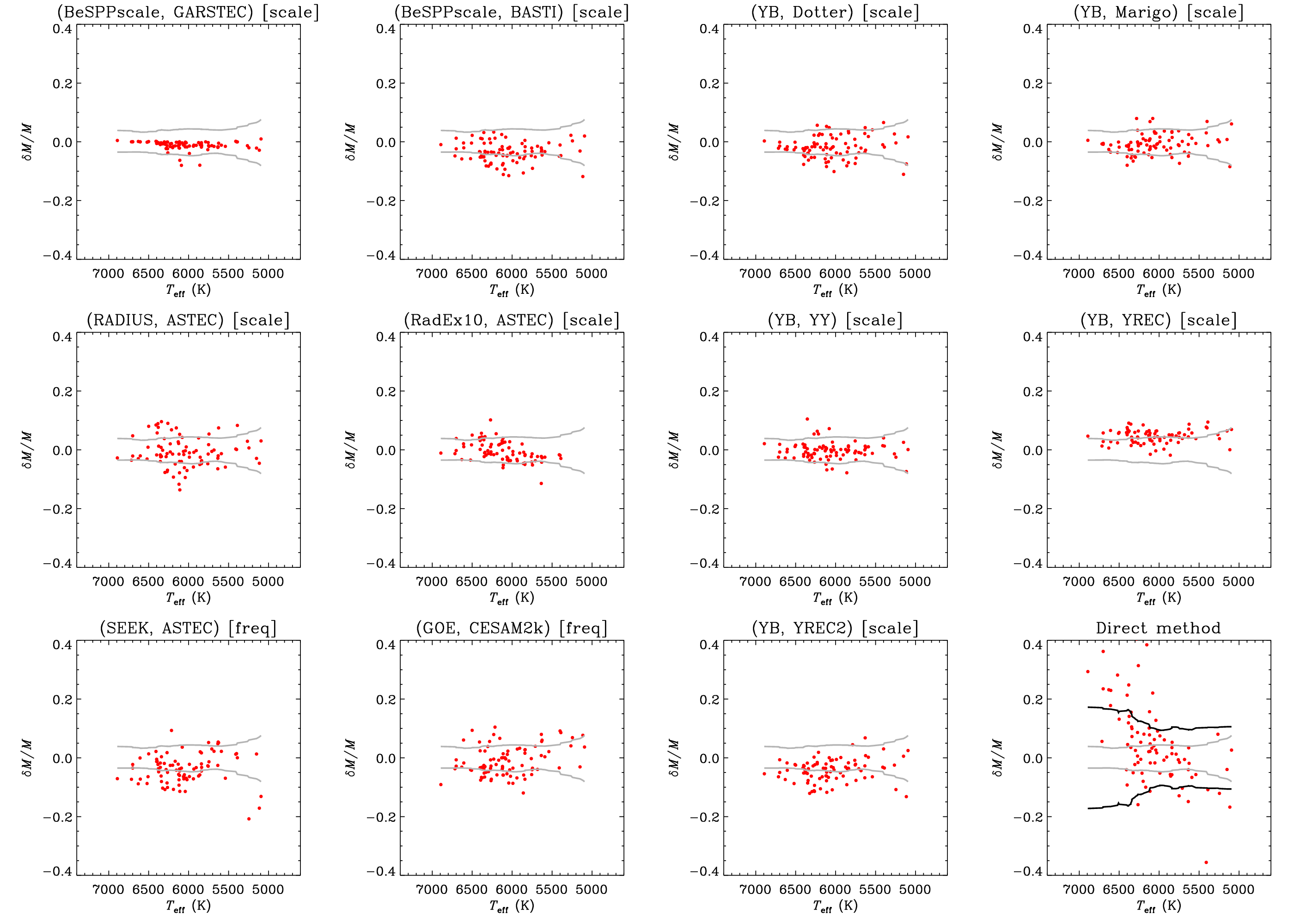}
\caption{As per Figure~\ref{fig:diffteff2_a}, but for fractional
  differences in mass $M$.}
\label{fig:diffteff2_b}
\end{figure*}

\begin{figure*}
\epsscale{1.0}
\plotone{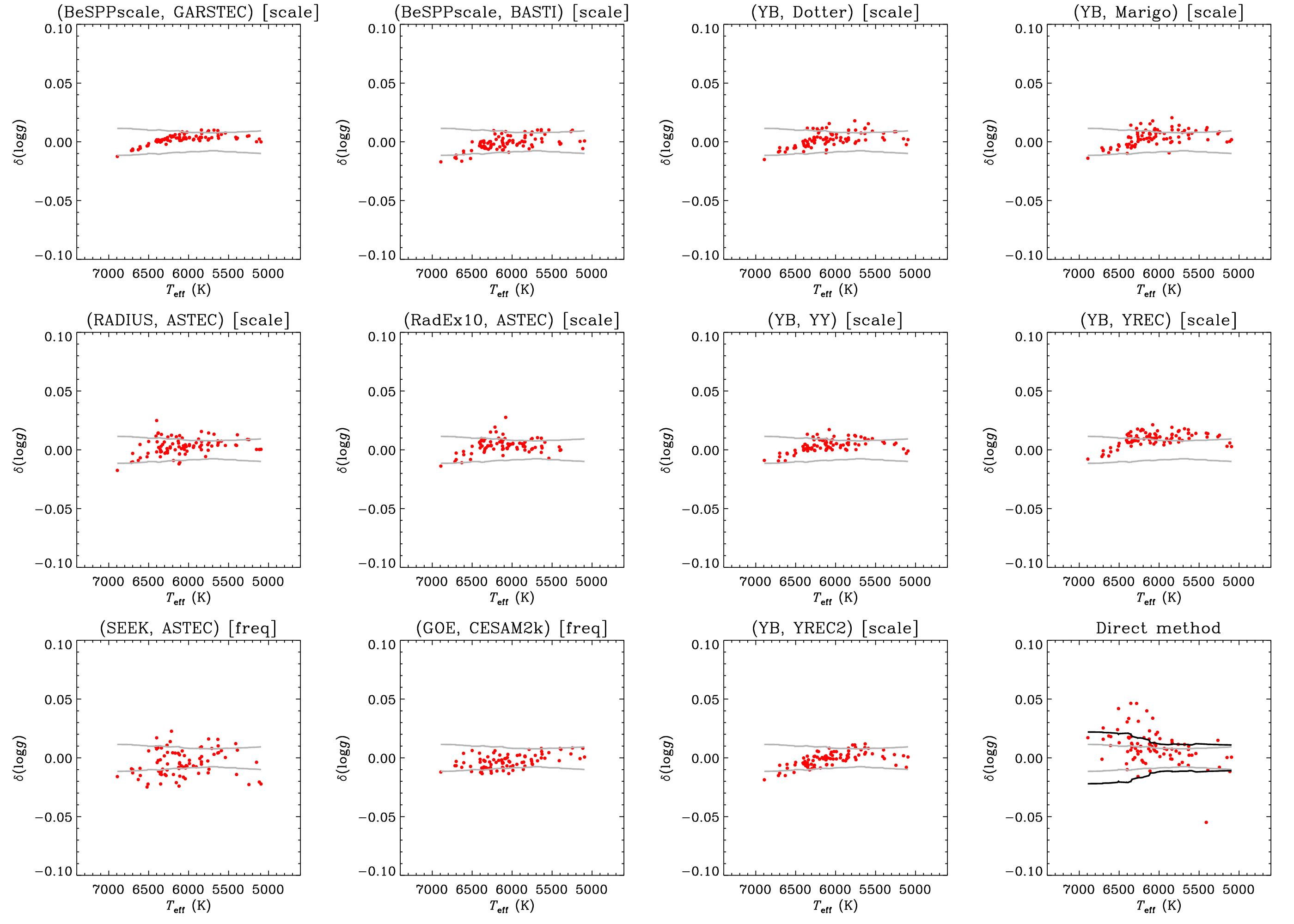}
\caption{As per Figure~\ref{fig:diffteff2_a}, but for differences in
  $\log\,g$.}
\label{fig:diffteff2_c}
\end{figure*}

\begin{figure*}
\epsscale{1.0}
\plotone{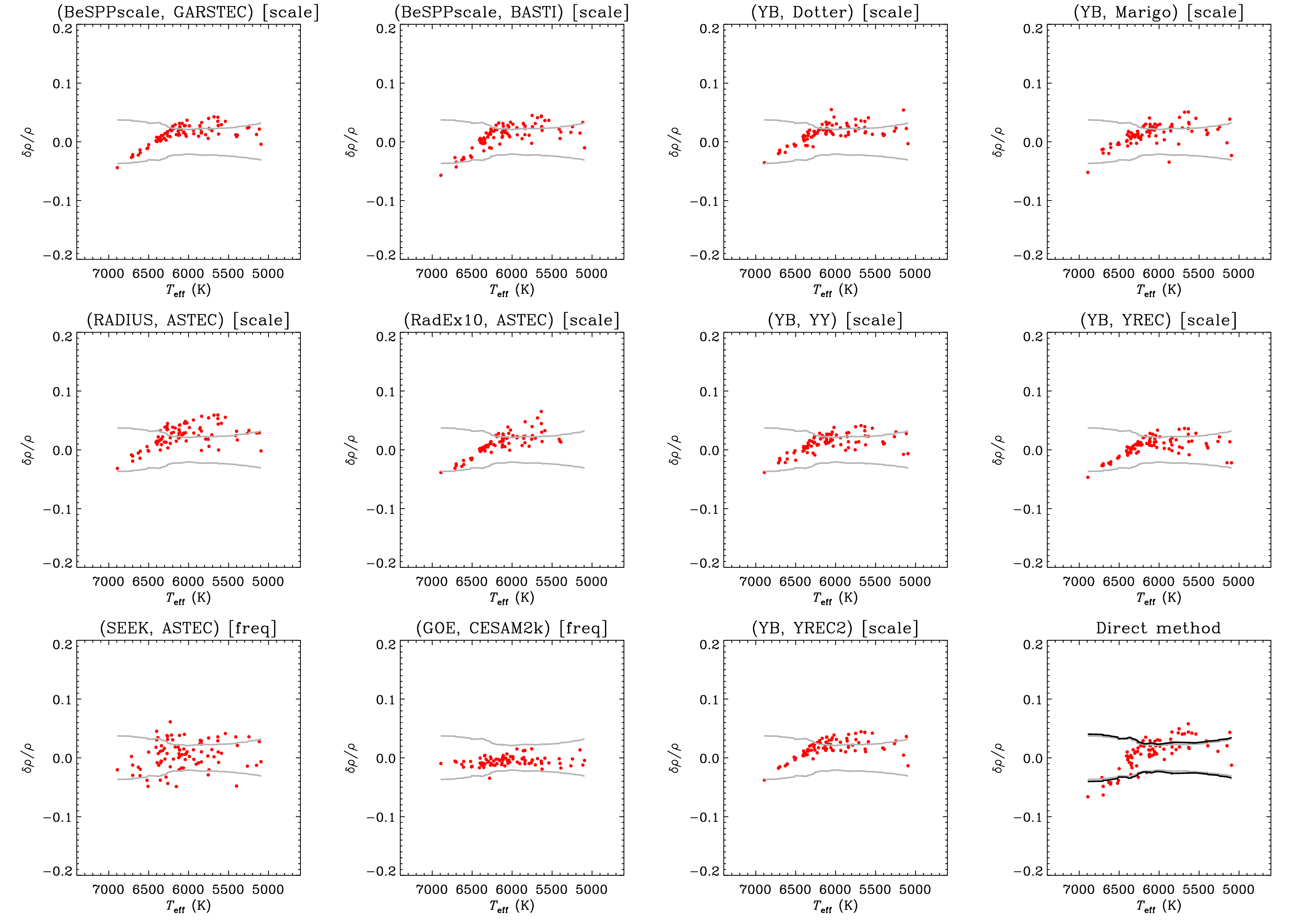}
\caption{As per Figure~\ref{fig:diffteff2_a}, but for fractional
  differences in average density $\rho$.}
\label{fig:diffteff2_d}
\end{figure*}

\begin{figure*}
\epsscale{1.0}
\plotone{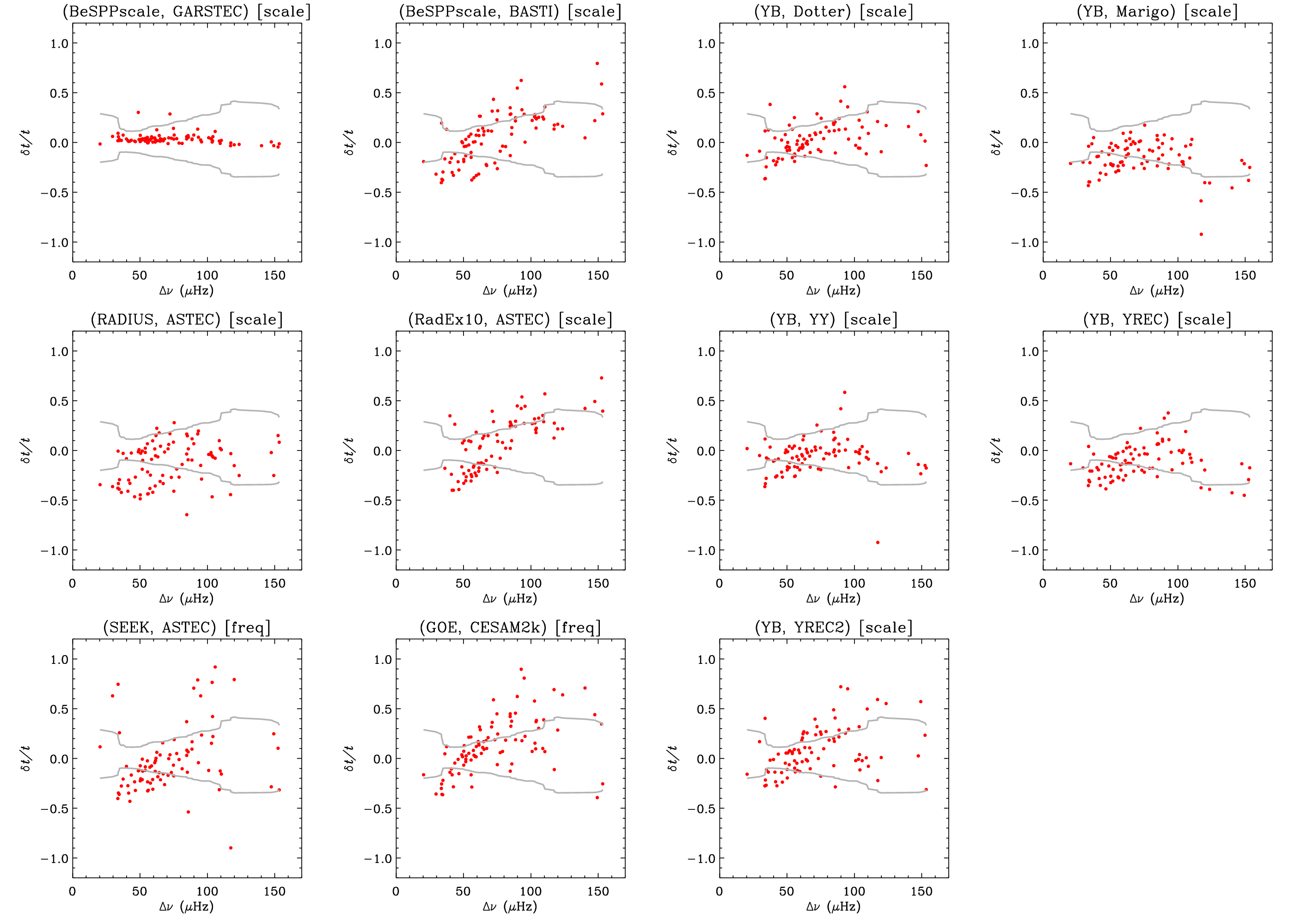}
\caption{As per Figure~\ref{fig:diffteff2_a}, but for fractional
  differences in age, $t$, and with \dnu\ as the
  independent variable for the plot.}
\label{fig:diffteff2_e}
\end{figure*}

Figures~\ref{fig:diffteff2_a} through~\ref{fig:diffteff2_e} show similar
visual comparisons of results, this time for analyses performed on the
cohort that had spectroscopic $T_{\rm eff}$ and [Fe/H] available.

Inspection of the results shows that in most cases the plotted
differences lie within the typical, median $1\sigma$ uncertainty
envelope. We see clearly the impact of adopting the scaling relations
to compute \dnu\, i.e., as manifested in the boomerang-shaped trends
which are most notable in the density plots, but also present in
radius, gravity and mass. These trends are noticeably lacking when
pipelines that used calculated eigenfrequencies are compared (e.g.,
see the flat density difference for the GOE pipeline, in
Figure~\ref{fig:diffteff1_d}. We note that, at the level of precision of
these data, the boomerang-shaped differences lie largely within the
median $1\sigma$ envelopes.

There is also evidence of model dependence in the results. For
example, masses and radii returned by GOE are both offset positively,
by different fractional amounts, relative to the reference BeSPP
results. These offsets combine to give a small negative, albeit flat
(see above) offset in density. Also noteworthy is the larger internal
scatter shown by the SEEK results (all properties), relative to the
internal scatter shown by the other pipelines.  SEEK is coupled to a
grid with multiple values of the mixing length parameter, $\alpha$.
However, not all subgrids constructed with a given $\alpha$ cover the
full range of metallicity. The increased scatter in the SEEK estimates
is a result of this non-uniform sampling of the $\alpha$-[Fe/H] space.
It is also worth remarking again on the scatter shown by the age
estimates.

%%%%%%%%%%%%%%%%%%%%%%%%%%%%%%%%%%%%%%%%%%%%%%%%%%%%%%%%%%%%%%%%%%%%%%%

% Fig. 13

% Fig. 14

% Fig. 15

% Fig. 16

% Fig. 17

%%%%%%%%%%%%%%%%%%%%%%%%%%%%%%%%%%%%%%%%%%%%%%%%%%%%%%%%%%%%%%%%%%%%%%%

% Fig. 18

% Fig. 19

% Fig. 20

% Fig. 21

% Fig. 22

%%%%%%%%%%%%%%%%%%%%%%%%%%%%%%%%%%%%%%%%%%%%%%%%%%%%%%%%%%%%%%%%%%%%%%%

\end{document}